\documentclass[aps,prb,twocolumn,groupedaddress,showpacs]{revtex4-1}

\usepackage{graphicx}
\usepackage{amsmath}

\begin{document}

\title{Thermodynamics of ultrasmall metallic grains in the presence of pairing and exchange correlations: mesoscopic fluctuations}

\author{Konstantin N. Nesterov}
\author{Y. Alhassid}
\affiliation{Center for Theoretical Physics, Sloane Physics Laboratory, Yale University, New Haven, Connecticut 06520, USA}

\date{\today}

\begin{abstract}
We study the mesoscopic fluctuations of thermodynamic observables in a nanosized metallic grain in which the single-particle dynamics are chaotic and the dimensionless Thouless conductance is large. Such a grain is modeled by the universal Hamiltonian describing the competition between exchange and pairing correlations.
The exchange term is taken into account exactly by a spin-projection method, and the pairing term is treated in the static-path approximation together with small-amplitude quantal fluctuations around each static fluctuation of the pairing field. Odd-even particle-number effects induced by pairing correlations are included using a number-parity projection. We find that the exchange interaction shifts the number-parity effects in the heat capacity and spin susceptibility to lower temperatures. In the regime where the pairing gap is similar to or smaller than the single-particle mean level spacing, these number-parity effects are suppressed by exchange correlations, and the fluctuations of the spin susceptibility may be particularly large. However, for larger values of the pairing gap, the number-parity effects may be enhanced by exchange correlations.
\end{abstract}


\maketitle

\newcommand{\bS}{{\hat{\bf S}}}
\newcommand{\hP}{{\hat{P}}}
\newcommand{\Tr}{{\mathrm{Tr}}}
\newcommand{\Th}{{\mathrm{Th}}}
\newcommand{\HBCS}{\hat{H}_{\mathrm{BCS}}}
\newcommand{\Dstatic}{{|\Delta_{0}|}}
\newcommand{\Dstaticarg}{{\Delta_0}}
\newcommand{\Auxfield}{{\widetilde{\Delta}}}
\newcommand{\Hstatic}{{\hat{H}_\Dstaticarg}}
\newcommand{\USPA}{\hat{U}_\Dstaticarg}
\newcommand{\Vtimedep}{\hat{V}}
\newcommand{\Vtimedepint}{\hat{V}_{\mathrm{int}}}

\setcounter{topnumber}{1} 

\section{Introduction\label{introduction}}

The physics of nano-sized metallic grains has attracted much attention following a series of experiments by Ralph, Black and Tinkham,~\cite{Ralph1995,Black1996,Ralph1997} in which individual energy levels of ultra-small aluminum grains were resolved by single-electron-tunneling spectroscopy.  Recent advances have made it possible to achieve better control over the size of the grain, which is important for quantitative measurements. In the experiments of Ref.~\onlinecite{Kuemmeth2008}, very high-quality spectra of chemically synthesized gold nano-particles were obtained.

Typical grains used in the spectroscopic experiments are in the ballistic regime, i.e., their size is smaller than the mean free path, and electron transport is determined by scattering from the boundaries of the grain rather than from impurities. When the boundaries are sufficiently irregular, the single-particle dynamics are chaotic. This induces  sample-specific fluctuations of observables, and the meaningful quantities are the statistical distributions of these observables; see Ref.~\onlinecite{Alhassid2000} and references therein.   The single-particle energies and wave functions follow the statistics of the random-matrix theory (RMT)~\cite{Mehta1991} in a Thouless energy window $E_\Th$ around the Fermi energy, where $E_\Th$ is determined by the time it takes for an electron to move across the grain.

When $E_\Th$ is much larger than the single-particle mean level spacing $\delta$, the grain is described by the so-called universal Hamiltonian.~\cite{Kurland2000,Aleiner2002} This Hamiltonian contains three interaction terms: a ``classical'' charging energy term, a pairing term that is characterized by a bulk pairing gap $\Delta$,  and exchange term that depends on the total spin of the grain and is characterized by a coupling constant $J_s$. These three interaction terms are universal, i.e., they are independent of the particular realization of the single-particle Hamiltonian. Here we assume $J_s/\delta < 1$ so that we are below Stoner instability of macroscopic polarization. 

When the pairing term is suppressed, (i.e., when only charging and exchange terms contribute), thermodynamic observables of the universal Hamiltonian can be calculated in closed form using a spin-projection method.~\cite{Alhassid2003} In Refs.~\onlinecite{Burmistrov2010} and \onlinecite{Burmistrov2012}, a Hubbard-Stratonovich transformation~\cite{Stratonovich1957, Hubbard1959} was employed to calculate in closed form observables such as the tunneling density of states and spin susceptibility.

In the absence of the exchange term, the universal Hamiltonian has the form of the Bardeen-Cooper-Schrieffer (BCS)~\cite{Bardeen1957} Hamiltonian. In the bulk limit $\Delta/\delta \gg 1$, an attractive pairing interaction leads to superconductivity. Effects of the BCS interaction in nano-sized metallic grains were studied extensively; see Ref.~\onlinecite{Vondelft2001} and references therein. Anderson argued~\cite{Anderson1959} that the smallest possible size of a system that can be a superconductor is determined by the condition $\Delta/\delta \sim 1$. In the experiments of Ref.~\onlinecite{Ralph1995,Black1996,Ralph1997}, a pairing gap was clearly observed in the excitation spectra of the largest aluminum grains containing an even number of electrons, while it was impossible to resolve such a gap in the smaller grains. This, however, does not necessarily mean that pairing correlations disappear in the smaller grains. It was proposed that thermodynamic properties could be a more suitable tool to probe this fluctuation-dominated regime, in which $\Delta/\delta \lesssim 1$.~\cite{DiLorenzo2000} Signatures of pairing correlations in this regime are the dependence of observables on the number parity of electrons in the grain. A good example is the re-entrant behavior (i.e., a local minimum) of the spin susceptibility with decreasing temperature in an odd grain.~\cite{DiLorenzo2000,Falci2000}  Odd-even effects in the heat capacity and magnetic susceptibility were experimentally observed in small palladium clusters.~\cite{Volokitin1996}

BCS theory breaks down when $\Delta/\delta \lesssim 1$ and fluctuations of the gap order parameter (beyond its mean-field BCS value) are important. In the static-path approximation (SPA),~\cite{Muhlschlegel1972, Alhassid1984, Lauritzen1988} only static fluctuations of the gap are taken into account.  A better approximation, the SPA plus random-phase approximation (RPA), takes into account small-amplitude time-dependent quantal fluctuations of the order parameter around each static field.~\cite{Kerman1981, Kerman1983, Puddu1991, Lauritzen1993, Rossignoli1997, Attias1997} Number-parity effects can be studied by using an exact number-parity projection.\cite{Goodman1981,Rossignoli1998,Balian1999}  The heat capacity and spin susceptibility of a metallic grain (without exchange correlations) as functions of temperature were studied in the SPA+RPA method together with a number-parity projection in Ref.~\onlinecite{Falci2002} as well as by quantum Monte Carlo methods~\cite{vanHoucke2006, Alhassid2007} and by Richardson's solution.~\cite{Richardson1963, Richardson1967,DiLorenzo2000,Alhassid2007} In all of those calculations, number-parity effects were clearly identified in both the heat capacity and spin susceptibility of the grain. Signatures of pairing correlations were also found in the spin susceptibility as a function of magnetic field.~\cite{Schechter2001, Gladilin2004}

The exchange interaction competes with the BCS-like pairing interaction.  Exchange tends to maximize spin polarization, while pairing correlations tend to minimize the spin.  It is known that, depending on the values of $\Delta/\delta$ and $J_s/\delta$, the ground state of a system can be superconducting, ferromagnetic, or one in which pairing and ferromagnetic correlations  coexist.~\cite{Ying2006, Schmidt2007} The effects of mesoscopic fluctuations on this competition were studied in Ref.~\onlinecite{Falci2003}.

The effect of both pairing and exchange correlations  on the thermodynamic properties of the grain (heat capacity and spin susceptibility) was studied in Ref.~\onlinecite{VanHoucke2010} for the case of an equally spaced single-particle spectrum by using a quantum Monte Carlo method.  These thermodynamic quantities can also be calculated directly from the eigenvalues of the universal Hamiltonian using Richardson's solution, modified to take into account the exchange interaction.~\cite{Schmidt2007} The combined effect of exchange and pairing interactions on the spin susceptibility as a function of magnetic field at zero temperature was studied in Ref.~\onlinecite{Schechter2004}.

In this work, we study the general problem of mesoscopic fluctuations of thermodynamic properties of the grain in the presence of both pairing and exchange correlations assuming spin-orbit coupling is negligible. The quantum Monte Carlo method and Richardson's solution mentioned above are computationally intensive, and are less practical in calculating the mesoscopic fluctuations of thermodynamic properties for which many realizations of the grain must be studied. Richardson's solution also becomes less tractable at larger values of the pairing gap or at higher temperatures, where a very large number of energy eigenvalues is required.

Here we use a more efficient method to calculate the heat capacity and spin susceptibility of the grain at finite temperature. The exchange interaction is treated exactly using a spin-projection method,~\cite{Alhassid2003, Alhassid2007a} and the corresponding spin-projected partition functions are calculated in the SPA+RPA approach. Number-parity effects are captured by a number-parity projection. This approach is particularly suitable for studying the mesoscopic fluctuations.

The outline of the paper is as follows. In Sec.~\ref{section_model}, we discuss the universal Hamiltonian and briefly review the conditions of its validity. In Sec.~\ref{section_theory}, we discuss the calculation of the canonical partition function, and use it to evaluate the heat capacity and spin susceptibility of a system described by the universal Hamiltonian. We also discuss the stability of the RPA, which is unstable below a certain critical value of the temperature. In Sec.~\ref{section_results}, we present our results and discuss their physical significance. We conclude in Sec.~\ref{section_conclusions}.

\section{Model \label{section_model}}

In a chaotic grain, the statistical fluctuations of single-particle energies and wave functions follow RMT~\cite{Mehta1991} at energy scales below $E_\Th$. In the absence of spin-orbit scattering and orbital magnetic field, the relevant ensemble is the Gaussian orthogonal ensemble (GOE). In general, the matrix elements of the electron-electron interaction in the basis of eigenstates of the non-interacting part of the Hamiltonian have a complex structure that depends on the particular realization of the one-body Hamiltonian. In the limit of a large Thouless conductance $g_\Th=E_\Th/\delta \gg 1$, these matrix elements can be decomposed into an average and fluctuating parts.~\cite{Kurland2000,Aleiner2002} The average interaction terms together with the one-body Hamiltonian are referred to as the universal Hamiltonian.~\cite{Kurland2000,Aleiner2002} The fluctuating (non-universal) part of the interaction forms an induced two-body ensemble~\cite{Alhassid2005_PRB} whose matrix elements are suppressed by $1/g_\Th$.

For a fixed number of electrons, the universal Hamiltonian has the form
\begin{equation}\label{universal_hamiltonian}
\hat{H} = \sum_{i, \sigma=\uparrow,\downarrow} \epsilon_i c^\dagger_{i\sigma}
c_{i\sigma} - g \hP^\dagger\hP - J_s \bS^2\,,
\end{equation}
where
\begin{equation}\label{pairing operators P+ and P}
\hP^\dagger = \sum_i c^\dagger_{i\uparrow}c^\dagger_{i\downarrow}\,,\,\,\,\, \hP =
\sum_i c_{i\downarrow} c_{i\uparrow}\,,
\end{equation}
and $\hat{\bf{S}}$ is the total spin. The single-particle levels are distributed as the eigenvalues of a GOE random matrix with a mean single-particle level spacing $\delta$. The universal interaction terms are invariant under orthogonal transformations of the single-particle basis, allowing us to write the one-body part in a diagonal form. If the particle number $N$ is not fixed, then the charging term $E_c\hat{N}^2$ should also be included in Eq.~(\ref{universal_hamiltonian}).  The bandwidth of the model $\sim  N_{\rm sp} \delta$, where $N_{\rm sp}$ is the number of single-particle levels, should satisfy the conditions $N_{\rm sp} \delta \gg \Delta$ and $N_{\rm sp} \delta \gg k T$ at temperature $T$ (where $k$ is the Boltzmann constant).

The condition of a large Thouless conductance $g_\Th \gg 1$ guarantees that the number of single-particle levels that follow RMT statistics within a Thouless energy window around the Fermi energy is sufficiently large, and that the  nonuniversal correction to the interaction can be ignored. We assume that the Thouless energy is larger than the bandwidth so that RMT is applicable in the full model space. This implies the conditions $E_\Th \gg kT$ and  $E_\Th \gg \Delta$. For ballistic grains, the latter assumption is equivalent to $L\ll \xi_0$, where $L$ is the linear size of the grain and $\xi_0$ is the superconducting coherence length. This condition also allows us to ignore the spatial fluctuations of the gap $\Delta$ within the grain, i.e.,  to treat the grain as a zero-dimensional object with respect to the fluctuations of the order parameter.

The pairing constant $g$ in the universal Hamiltonian depends on  the number of single-particle orbitals in the model space. To reduce the computational effort, we choose  $N_{\rm sp}$ to be smaller than the number of orbitals within the physical window of the Debye frequency around the Fermi energy. In doing so we renormalize the value of $g$ according to~\cite{Berger1998,Alhassid2007}
\begin{equation}
   \frac{g}{\delta} =
    \frac{1}{\mathrm{arcsinh}\left(\frac{N_{\text{sp}}/2}{\Delta/\delta}\right)}\,,
\end{equation}
where we have taken the Fermi level to be in the middle of the single-particle spectrum (i.e., we assume half-filling). Depending on the temperature, we choose $N_{\text{sp}}$ to be between 30 and 60 in our calculations to ensure the condition $N_{\text{sp}} \delta \gg kT$.  In our studies,  $\Delta/\delta \lesssim 5.0$ so the condition $N_{\text{sp}} \delta \gg \Delta$ is also satisfied.

To study the mesoscopic fluctuations of thermodynamic observables, we generate a large number ($\sim 1000$) of realizations of the single-particle spectrum in Eq.~(\ref{universal_hamiltonian}) and calculate these observables for each of them.

\section{Theory \label{section_theory}}

In this section, we present approximate analytical results for the partition function and the spin susceptibility of a system described by the Hamiltonian (\ref{universal_hamiltonian}) for a given realization of the single-particle spectrum.  First, we show that both quantities can be related to the $S_z$-projected partition function in the absence of exchange ($S_z$ is the spin projection along the $z$ axis). Second, we present an auxiliary-field path-integral formalism for treating the pairing interaction and explain how the integral is evaluated in the SPA+RPA method. Third, we discuss the inclusion of number-parity projection.

\subsection{Spin-projection method}

The universal Hamiltonian (\ref{universal_hamiltonian}) commutes with the
total-spin and particle-number operators. Consequently, the corresponding  partition function at temperature $T=1/\beta$ (here and in the following we set the Boltzmann constant $k=1$)  can be
written as  
\begin{multline}\label{Z^(J)_1}
    Z(J_s) = \Tr\, e^{-\beta\hat{H}} =\Tr
    \,\,e^{-\beta(\HBCS-J_s\hat{\bf{S}}^2)} \\
    = \sum_{S} e^{\beta
    J_s S(S+1)} \Tr_{S} \left(e^{-\beta\HBCS}\right)\,,
\end{multline}
where $\Tr_{S}$ is the trace restricted to states with fixed spin $S$, and
$\HBCS$ is the BCS-like pairing Hamiltonian
\begin{equation}\label{BCS_Hamiltonian}
    \hat{H}_{BCS} = \sum_{i,\sigma=\uparrow,\downarrow} \epsilon_i c^\dagger_{i\sigma}
    c_{i\sigma} - g \hP^\dagger\hP\,.
\end{equation}
Similarly, the spin susceptibility (at zero external Zeeman field) can be written as
\begin{multline}\label{chi_1}
\chi = 4\beta\mu_B^2\langle\hat{S}^2_z\rangle = \frac{4\beta\mu_B^2}{3}\langle{\bf\hat{S}}^2\rangle \\
= \frac{4\beta\mu_B^2}{3} \frac{\sum_S S(S+1) e^{\beta J_sS(S+1)}\mathrm{Tr}_{S} \left(e^{-\beta \HBCS}\right)}{Z(J_s)} \,,
\end{multline}
where $\hat{S}_z$ is the spin-component operator along the $z$ direction and $\mu_B$ is the Bohr magneton.

For a scalar operator (i.e., an operator that is rotational invariant in spin space) $\hat{X}$~\cite{Alhassid2003,Alhassid2007a}
\begin{equation}\label{Tr_S = Tr_Sz-Tr_Sz+1}
    \mathrm{Tr}_S \hat{X} = (2S+1) \left(\mathrm{Tr}_{S_z=S}\hat{X} -
    \mathrm{Tr}_{S_z=S+1}\hat{X}\right)\,,
\end{equation}
where $\mathrm{Tr}_{S_z = M}$ denotes the trace restricted to states
with a fixed spin component $S_z=M$ (while the spin quantum number is no longer fixed).
Using Eq.~(\ref{Tr_S = Tr_Sz-Tr_Sz+1}), we can express the partition function (\ref{Z^(J)_1}) and spin susceptibility (\ref{chi_1}) of a system described by the Hamiltonian (\ref{universal_hamiltonian}) in terms of the $S_z$-projected partition function of the corresponding system in the absence of exchange interaction as
\begin{equation}\label{Z^(J)}
 Z(J_s)  = \sum_S (2S+1) e^{\beta J_sS(S+1)}\left(Z_{S_z=S}-Z_{S_z=S+1}\right)\,,
\end{equation}
and
\begin{multline}\label{chi}
{\chi}=\frac{4\beta\mu_B^2}{3} \frac{1}{Z(J_s)}
\sum_S S(S+1)(2S+1) \\ \times e^{\beta J_sS(S+1)}\left(Z_{S_z=S}-Z_{S_z=S+1}\right)\,.
\end{multline}
Here
\begin{equation}\label{Z_NS_z}
 Z_{S_z=M} = \mathrm{Tr}_{S_z=M} \left(e^{-\beta \HBCS}\right)
\end{equation}
is the $S_z$-projected partition function of the BCS-like Hamiltonian (\ref{BCS_Hamiltonian}).

\subsection{Auxiliary-field path-integral formulation}

In the following we consider a partition function of the type
\begin{equation}\label{Z_lambda}
Z_\lambda = \Tr_\lambda\hat{U}\,,
\end{equation}
where $\hat U$ is the propagator of the BCS Hamiltonian
 \begin{equation}\label{BCS_propagator}
\hat{U} = e^{-\beta\left(\HBCS - \mu\hat{N}\right)}\,,
\end{equation}
and $\Tr_\lambda$ denotes a trace with certain restrictions (e.g., fixed $S_z$).  Using the auxiliary-field path-integral representation of the propagator $\hat U$, we discuss the static-path approximation (SPA) as well as the quantal RPA-like corrections around each static fluctuation.
We have included a $\mu \hat N$ term in the propagator (\ref{BCS_propagator}) because further calculations of the partition function (\ref{Z_NS_z}) will be carried out in the grand-canonical formalism.

 The interaction term in the BCS Hamiltonian (\ref{BCS_Hamiltonian}) can be decoupled by means of a Hubbard-Stratonovich
 transformation.~\cite{Stratonovich1957, Hubbard1959}  In this transformation, we express the propagator (\ref{BCS_propagator}) as a functional integral over a complex auxiliary field $\Auxfield(\tau)$ that depends on imaginary time $\tau$ ($0\leq\tau\leq\beta$) as follows~\cite{Puddu1991, Lauritzen1993, Rossignoli1998}
 \begin{equation}\label{HS_transformation}
\hat{U} = \int {\mathcal{D}}[\Auxfield,\Auxfield^*] e^{- \int\limits_0^\beta d\tau |\Auxfield(\tau)|^2/g} \hat{U}_\Auxfield\,.
\end{equation}
Here,
\begin{equation}\label{U_eff_aux_field}
\hat{U}_\Auxfield = {\cal T}e^{- \int\limits_0^\beta d\tau
\,\hat{H}_{\Auxfield(\tau)} }\,
\end{equation}
is the propagator for the one-body Hamiltonian
\begin{multline}\label{H_eff(xi_1,xi_2)}
\hat{H}_\Auxfield
=\sum_i \left[\left(\epsilon_i - \mu-\frac
g2\right)\left(c^\dagger_{i\downarrow}c_{i\downarrow}+c^\dagger_{i\uparrow}c_{i\uparrow}\right)\right. \\ \left. - \Auxfield\, c^\dagger_{i\uparrow} c^\dagger_{i\downarrow} - \Auxfield^* \,c_{i\downarrow}c_{i\uparrow}+
\frac{g}{2}\right]\,
\end{multline}
and ${\cal T}$ denotes time ordering. The measure ${\mathcal{D}}[\Auxfield,\Auxfield^*]$ is defined to  preserve the  normalization (i.e., $\hat{U} = 1$ when $\hat{U}_\Auxfield \equiv 1$)
\begin{equation}
{\mathcal{D}}[\Auxfield,\Auxfield^*]= \lim_{M\rightarrow \infty}
\prod_{m=1}^M \int \frac{\Delta\beta}{2\pi g}\,\, d\Auxfield_{m}d\Auxfield^*_m \,,
\end{equation}
 where $\Delta \beta = \frac{\beta}{M}$.

The auxiliary field $\tilde\Delta(\tau)$ can be separated into its static and $\tau$-dependent parts by a Fourier series
\begin{equation}
	\Auxfield(\tau) = \Delta_0 + \sum_{r\ne 0} \Delta_r e^{i\omega_r\tau}\,,
\end{equation}
where  $\omega_r = {2\pi r}/{\beta}$ ($r$ integer) are the bosonic Matsubara frequencies. In the SPA,  $\Auxfield(\tau)$ is replaced by $\Dstaticarg$ in the Hamiltonian (\ref{H_eff(xi_1,xi_2)}), and the propagator (\ref{U_eff_aux_field}) is approximated by
\begin{equation}\label{U_eff_static}
\hat{U}_\Auxfield \approx \USPA = e^{-\beta \Hstatic }\,,
\end{equation}
where
$\Hstatic$ is the Hamiltonian (\ref{H_eff(xi_1,xi_2)}) for a static field $\Auxfield=\Dstaticarg$. The BCS theory can be derived by applying the saddle-point approximation to the SPA integral.

The time-dependent part of Eq.~(\ref{H_eff(xi_1,xi_2)}) can be written as
\begin{equation}
\Vtimedep(\tau)= - \sum_{i,r\neq 0} e^{i\omega_r
\tau}\left(\Delta_r \,c^\dagger_{i\uparrow}
c^\dagger_{i\downarrow} + \Delta^*_{-r} \,c_{i\downarrow}c_{i\uparrow} \right) \,.
\end{equation}
Since we are interested in the correction to the SPA propagator (\ref{U_eff_static}), it is natural to use an interaction representation of $\Vtimedep(\tau)$ with respect to the unperturbed static Hamiltonian $\Hstatic$
\begin{equation}\label{V_intsigma}
\Vtimedepint(\tau) = e^{\tau \Hstatic}
\Vtimedep(\tau)e^{-\tau \Hstatic}\,.
\end{equation}

To obtain the expression for $\Vtimedepint$, it is convenient to work in the $\Dstaticarg$-dependent basis that diagonalizes the static Hamiltonian $\Hstatic$. We will refer to this basis as the quasiparticle representation (even though this is strictly the case only for the saddle-point value of $\Dstaticarg$).
For a given value of $\Dstaticarg$, the quasiparticle operators are related to the original particle operators via the  Bogoliubov transformation
\begin{equation}\label{Bogoliubov_transf}
\left\{
\begin{array}{c}
c_{i\uparrow} = u_i a_{i\uparrow}+v_i \,e^{i\theta}\,a^\dagger_{i\downarrow}\,, \\
c_{i\downarrow} = u_i a_{i\downarrow} - v_i \,e^{i\theta}\, a_{i\uparrow}^\dagger\,.
\end{array}
\right.
\end{equation}
Here
\begin{equation}
u_i^2 = \frac 12 (1+\gamma_i)\,, \quad
v_i^2 = \frac 12 (1-\gamma_i)\,, \quad
\theta =\text{arg} \,\Dstaticarg \,,
\end{equation}
where
\begin{equation}\label{gamma_i}
 \gamma_i = \frac{\epsilon_i - \mu - \frac g2}{E_i}\,,
\end{equation}
and $E_i$ are the quasiparticle energies
\begin{equation}\label{E_i}
 E_i =
\sqrt{\left(\epsilon_i - \mu - \frac g2\right)^2 + \Dstatic^2} \;.
\end{equation}
 In this quasiparticle basis, $\Hstatic$ and $\Vtimedep(\tau)$ have the  forms
\begin{equation}\label{H_0_qp_basis}
\Hstatic = \sum_i \left[\epsilon_i-\mu -
E_i \,+\, E_i \left(a^\dagger_{i\uparrow}
a_{i\uparrow} + a^\dagger_{i\downarrow}
a_{i\downarrow}\right)\right]
\end{equation}
and
\begin{multline}\label{V_sigma_quasiparticle_basis}
\Vtimedep(\tau) = - \frac 12 \sum_{i, r \neq 0 }  \, e^{i\omega_r \tau} \left[\vphantom{\frac{\Dstatic}{E_i}}
\Lambda_{ir}\, a^\dagger_{i\uparrow} a^\dagger_{i\downarrow}  +
\Lambda^*_{i,-r}\, a_{i\downarrow} a_{i\uparrow} \right. \\ \left. +
\frac{\Dstatic}{E_i} \left(\Delta_re^{-i\theta}+\Delta^*_{-r}e^{i\theta}\right)\left(1 - a^\dagger_{i\downarrow}
a_{i\downarrow} - a^\dagger_{i\uparrow}a_{i\uparrow}\right)\right] \,
\end{multline}
with
\begin{equation}\label{Lambda_ir}
\Lambda_{ir} = (\gamma_i+1)\Delta_r + (\gamma_i-1)e^{2i\theta}\Delta^*_{-r}\,.
\end{equation}
The dependence of $\Vtimedep(\tau)$ on $\theta$ can be eliminated by an appropriate gauge transformation of $\Delta_r$ and the quasiparticle operators. Therefore, the integrand in the SPA integral depends only on the absolute value of  $\Dstaticarg$.

The form of $\Vtimedepint(\tau)$ in the quasiparticle basis is the same as of $\Vtimedep(\tau)$ in Eq.~(\ref{V_sigma_quasiparticle_basis}) with $a$ and $a^\dagger$ replaced by
\begin{equation}\label{a(tau)=a*e}
\begin{array}{l}
 a_{i\sigma}(\tau) = a_{i\sigma}e^{-E_i\tau}\,,\,\,\,\,
a^\dagger_{i\sigma}(\tau) = a^\dagger_{i\sigma}e^{E_i\tau}\,.
 \end{array}
\end{equation}

In the interaction representation,
\begin{equation}
\hat{U}_\Auxfield  = \USPA \,\,\hat{\mathcal{U}}_{\mathrm{int}}\,,
\end{equation}
where
\begin{equation}
\hat{\mathcal{U}}_{\mathrm{int}} = {\cal T}e^{-\int\limits_0^\beta d\tau \Vtimedepint(\tau)}\,.
\end{equation}
As a result, the partition function (\ref{Z_lambda}) can be written as
\begin{equation}\label{Z_lambda_func_int}
Z_\lambda =\frac{\beta}{g} \int \limits_0^\infty  \,d\,\Dstatic^2 e^{-{\beta\over g}\Dstatic^2} \,\,Z_{\lambda}(\Dstaticarg) \,\,C_\lambda(\Dstaticarg)\,,
\end{equation}
where
\begin{equation}\label{Z^SPA}
Z_{\lambda}(\Dstaticarg) = \Tr_\lambda \USPA\,
\end{equation}
is the partition function for a static value $\Dstaticarg$ of the pairing field, and
\begin{equation}\label{C_lambda}
C_\lambda(\Dstaticarg) = \int \mathcal{D}'[\Delta_r,\Delta^*_r]\, e^{-(\beta/g) \sum_{r\ne 0}|\Delta_r|^2}\,\, \left\langle \hat{\mathcal{U}}_{\mathrm{int}}\right\rangle_{\lambda, \Dstaticarg}\,
\end{equation}
is the quantum correction.
The average in Eq.~(\ref{C_lambda}) is defined with respect to the static-field  propagator
\begin{equation} \label{average_lambda}
\left\langle \hat{A} \right\rangle_{\lambda,\Dstaticarg} = \frac{\Tr_\lambda\left(\USPA \,\,\hat{A}\right)}{\Tr_\lambda \USPA}\,,
\end{equation}
and the integration measure is
\begin{equation}
\mathcal{D}'[\Delta_r,\Delta^*_r] = \prod_{r\ne 0}\frac{\beta d\Delta_r d\Delta^*_r}{2\pi g}\,.
\end{equation}

Equation (\ref{Z_lambda_func_int}) for the partition function $Z_\lambda$ is exact. In the SPA, $C_\lambda = 1$ or, equivalently, $\Vtimedep (\tau) = 0$. The SPA+RPA approximation is obtained by evaluating the integral over $\Delta_r$ and $\Delta^*_r$ in Eq.~(\ref{C_lambda}) in the saddle-point approximation assuming small-amplitude fluctuations. To this end, we write
\begin{equation}\label{cumulant_exp}
\ln\left \langle \hat{\mathcal{U}}_{\mathrm{int}} \right\rangle_{\lambda,\Dstaticarg} \approx  \frac 12 \int_0^\beta \int_0^\beta d\tau d\tau' \langle {\cal T}\,\Vtimedepint(\tau) \Vtimedepint(\tau') \rangle_{\lambda,\Dstaticarg}\,,
\end{equation}
where we have assumed that $\int_0^\beta d\tau \langle \Vtimedepint(\tau)\rangle_{\lambda,\Dstaticarg} = 0$. This is valid provided the projection in  $\Tr_\lambda$ conserves the quasiparticle occupation number, which is the case for the $S_z$-projected trace but not for the canonical (particle-projected) trace. The RPA correction factor (\ref{C_lambda}) is then given by
\begin{multline}\label{C_RPA_definition}
C_\lambda^{\mathrm{RPA}}(\Dstaticarg) = \int \mathcal{D}'[\Delta_r,\Delta^*_r]\, e^{-{\beta\over g} \sum_{r\ne 0}|\Delta_r|^2}\\ \times\exp\left(\frac 12  \int_0^\beta \int_0^\beta d\tau d\tau' \left\langle {\cal T}\,\Vtimedepint(\tau) \Vtimedepint(\tau') \right\rangle_{\lambda,\Dstaticarg}\right)\,.
\end{multline}
Higher-order corrections can be obtained by including more terms in the cumulant expansion (\ref{cumulant_exp}).

\subsection{Canonical and number-parity projections}

The canonical partition function can be related to the grand-canonical partition by particle-number projection.~\cite{Alhassid2000} However, an exact particle-number projection ``inside'' the integral (\ref{Z_lambda_func_int})  for each value of the static field $\Dstaticarg$ leads to a complicated expression since the particle-number operator $\hat{N}$ does not commute with the static Hamiltonian $\Hstatic$.~\cite{Note1}

In the following we will carry out the particle-number projection in the saddle-point approximation. However, to account for important odd-even effects, we project on the number parity of electrons. The trace in (\ref{Z_lambda}) will be restricted to $S_z$ and the number-parity quantum number $\eta$  ($\eta=+1$ for an even number of particles, and $\eta=-1$ for an odd number), i.e., $\lambda = \eta, S_z$.

The canonical partition function for $N$ particles can be obtained from Eq.~(\ref{Z_lambda}) by particle-number projection
\begin{multline}\label{mu-integral}
 Z_{N,\lambda} = \int\limits_{-\pi}^{\pi} \frac{d\alpha}{2\pi} e^{-i\alpha N} \,\Tr_\lambda \left(e^{-\beta \HBCS + i\alpha \hat{N}}\right)\\
 =\int\limits_{-i\pi/\beta}^{i\pi/\beta}\frac{\beta \,d\mu}{2\pi i} \int \limits_0^\infty \frac{\beta \,d\,\Dstatic^2}{g} e^{-\beta [F_\lambda(\mu,\Dstaticarg)+\mu N]}C_\lambda(\Delta_0)\,,
 \end{multline}
 where
\begin{equation}\label{F_lambda}
e^{-\beta F_\lambda(\mu,\Dstaticarg)} = e^{-(\beta/g)\Dstatic^2} \,\,Z_{\lambda}(\Dstaticarg)\,.
\end{equation}
Here $Z_{\lambda}(\Dstaticarg)$ and $C_\lambda(\Delta_0)$ are calculated for a chemical potential $\mu$. We denote by $F$ the grand-canonical free energy (without the restriction $\lambda$) given by
\begin{multline}
e^{-\beta F(\mu,\Dstaticarg)} = e^{-(\beta/g)\Dstatic^2} \Tr e^{-\beta \Hstatic} \\
 = e^{-(\beta/g)\Dstatic^2}\prod_i 4e^{-\beta(\epsilon_i-\mu)} \cosh^2 \frac{\beta E_i}{2}\,.
\end{multline}
We exchange the order of integrations in Eq.~(\ref{mu-integral}) and for each value of $\Dstaticarg$ evaluate the integral over $\mu$ by the saddle-point approximation. The latter is applied to the function $e^{-\beta [F(\mu,\Dstaticarg)+\mu N]}$ considering the remaining part, $e^{-\beta(F_\lambda-F)}C_\lambda(\Dstaticarg)$, as a pre-factor that does not affect the saddle-point integration. We then obtain~\cite{Alhassid2005_PRC}
\begin{multline}
Z_{N,\lambda} \approx
\int \limits_0^\infty \frac{\beta \,d\,\Dstatic^2}{g} \\
\times \left(\frac{2\pi}{\beta}\left|\frac{\partial^2 F}{\partial \mu^2}\right|\right)^{-1/2} e^{-\beta [F_\lambda(\mu,\Dstaticarg)+\mu N]}C_\lambda(\Dstaticarg)\,,
\end{multline}
where
\begin{equation}\label{d2Fdmu2}
\frac{\partial^2 F}{\partial \mu^2}= - \sum_i\frac{\beta
E_i(\epsilon_i - \mu -\frac g2)^2 + \Dstatic^2\sinh(\beta E_i)
}{2E_i^3\cosh^2\left(\frac{\beta E_i}{2}\right)}\,,
\end{equation}
and $\mu$ is a $\Dstaticarg$-dependent chemical potential determined by
\begin{equation}\label{N=-dF/dmu}
N = -\frac{\partial F}{\partial \mu}= \sum_i\left(1 -
\frac{\epsilon_i - \mu - \frac g2}{E_i}\tanh\frac{\beta
E_i}{2}\right)\,.
\end{equation}

The number-parity projection is carried out using the projector
\begin{equation}\label{P_eta_definition}
 \hP_\eta = \frac 12 \left(1+\eta e^{i\pi \hat{N}}\right)\,,
\end{equation}
where $\eta = \pm 1$, depending on the number-parity of electrons. If this operator is inserted in the grand-canonical trace, only states with even (odd) number of particles will be taken into account for $\eta=1$ ($\eta = -1$).

\subsection{Number-parity and $S_z$-projected static partition function}

Here we discuss the calculation of the static partition function (\ref{Z^SPA}), when $\lambda$ corresponds to the projections on  number parity and $S_z$, i.e., $\lambda = \eta, S_z$.

The trace over states with fixed $S_z$ can be calculated exactly through a discrete Fourier transform as long as the maximal value of $S_z$ is finite. The corresponding projection operator is
\begin{equation}\label{P_S_z}
\hat{P}_{S_z} = \frac{1}{2S_{\rm max}+1}\sum_{m=-S_{\rm max}}^{S_{\rm max}} e^{i\phi_m (\hat{S}_z - S_z)} \,,
\end{equation}
where $S_{\rm max}$ is the maximal possible value of the spin and  $\phi_m = 2\pi m/(2S_{\rm max}+1)$ are quadrature points. We use this operator in accordance with the number-parity projection, i.e., the values of $m$ are integers or half-integers for even or odd number of electrons, respectively. Because our goal is to obtain an expression for the canonical partition function, the value of $S_{\rm max}$ is determined by the number of particles, rather than by the size of the Hilbert space.

The spin projection operator $\hat{S}_z$ commutes with  $\Hstatic$, so we can write
\begin{equation}
\mathrm{Tr}\left(\hat{P}_\eta e^{i\phi_m \hat{S}_z} e^{-\beta
 \Hstatic}\right) =  \mathrm{Tr}\left(\hat{P}_\eta  e^{-\beta
 \Hstatic+i\phi_m \hat{S}_z}\right) \;.
\end{equation}
 The second-quantized forms of $\hat{S}_z$ and $\hat{P}_\eta$ remain invariant under Bogoliubov transformation (\ref{Bogoliubov_transf}). Consequently, the projected partition function for a static pairing field $\Dstaticarg$ is given by
\begin{equation}\label{Z(Dstatic)}
Z_{\eta, S_z}(\Dstaticarg)=  \sum_{m}\frac{e^{-i\phi_m S_z}}{2S_{max}+1} \frac{Z^{(0,\phi_m)}(\Dstaticarg)+\eta \,Z^{(i\pi, \phi_m)}(\Dstaticarg)}{2}\,,
\end{equation}
where
\begin{multline}\label{Z_0 phi_m}
 Z^{(0,\phi_m)}(\Dstaticarg) = \mathrm{Tr}\left(e^{-\beta
 \Hstatic +i\phi_m \hat{S}_z}\right)
 \\
 = \prod_i e^{-\beta(\epsilon_i-\mu-E_i)} \left|1+e^{-\beta E_i - \frac{i\phi_m}{2}}\right|^2\,,
\end{multline}
and
\begin{multline}\label{Z_ipi phi_m}
Z^{(i\pi, \phi_m)}(\Dstaticarg) = \mathrm{Tr}\left(e^{i\pi\hat{N}}e^{-\beta
 \Hstatic +i\phi_m \hat{S}_z}\right)  
 \\
 =\prod_i e^{-\beta(\epsilon_i-\mu-E_i)} \left|1-e^{-\beta E_i - \frac{i\phi_m}{2}}\right|^2 \,.
\end{multline}
The last two expressions were obtained using the grand-canonical formalism.

\subsection{The RPA correction}\label{RPA_correction}

The RPA correction factor (\ref{C_RPA_definition}) is calculated in  Appendix \ref{Appendix_RPA_correction}.
For $\lambda=\eta, S_z$, it is given by
\begin{equation}\label{C_RPA = prod 1/det}
C^{\mathrm{RPA}}_{\lambda}(\Dstaticarg) = \prod_{r>0} \left[\det A(\omega_r)
\right]^{-1}\,,
\end{equation}
where
\begin{equation}\label{A-matrix}
A(\omega_r) = \left(%
\begin{array}{cc}
  1-g\sum_i\frac{2E_i\gamma_i^2f_{\lambda
i}}{4E_i^2+\omega_r^2} & g\sum_i\frac{\omega_r\gamma_i
f_{\lambda
i}}{4E_i^2+\omega^2_r} \\
  -g\sum_i\frac{\omega_r\gamma_i f_{\lambda
i}}{4E_i^2+\omega^2_r} & 1-g\sum_i\frac{2E_if_{\lambda
i}}{4E_i^2+\omega_r^2} \\
\end{array}%
\right)\,,
\end{equation}
and
\begin{equation}\label{F_eta_i}
f_{\lambda i} = \frac{1}{\beta} \frac{\partial \ln Z_{\lambda}(\Dstaticarg)}{\partial E_i}\,.
\end{equation}
Here $\omega_r=2\pi r/\beta$ are the positive bosonic Matsubara frequencies, and $\gamma_i$ are given by Eq.~(\ref{gamma_i}).  Each factor in the product (\ref{C_RPA = prod 1/det}) is obtained
 after a Gaussian integration over $\Delta_r,\Delta^*_r, \Delta_{-r}\, \text{and}\, \Delta^*_{-r}$. This integral converges if and only if the real parts of both eigenvalues of the matrix $A(\omega_r)$ are positive. Since  $A(\omega_r)$  is a $2\times 2$ real matrix, this is equivalent to
\begin{equation}\label{RPA_condition}
\mathrm{det}A(\omega_r) >0 \quad \text{and} \quad \mathrm{tr} A(\omega_r)>0 \,.
\end{equation}
For a given $\Dstaticarg$, the RPA breaks down below a certain temperature for which one of the conditions in Eq.~(\ref{RPA_condition}) is not satisfied.  It is then necessary to include higher-order cumulants in the expansion (\ref{cumulant_exp}). The highest temperature for which the RPA breaks down for at least one value of $\Dstaticarg$ is known as the SPA+RPA critical temperature $T_*$. The SPA+RPA approach is thus valid for temperatures above $T_*$. Numerical simulations show that $T_*$ increases with the BCS gap $\Delta$ and becomes of the order of $\delta$ for $\Delta/\delta \sim 5$. It has been recently proposed~\cite{Ribeiro2012} that the above instability can be avoided by treating non-perturbatively a low-energy collective mode.

In Appendix \ref{Appendix_RPA_matrix}, we show that the RPA correction factor (\ref{C_RPA = prod 1/det}) can  be written as
\begin{equation}\label{C_RPA = sinh/sinh}
C^{\mathrm{RPA}}_\lambda (\Dstaticarg) = \prod_i
\frac{\Omega_i}{2E_i}\frac{\sinh(\beta
E_i)}{\sinh\left(\frac{\beta \Omega_i}2\right)}\,,
\end{equation}
where $\pm \Omega_i$ are the eigenvalues of
the $2N_{\text{sp}}\times 2N_{\text{sp}}$ RPA matrix ($N_{\text{sp}}$ is the number of single-particle orbitals)
\begin{equation}\label{RPA_matrix}
\left(%
\begin{array}{cc}
  2E_i \delta_{ij} - \frac g2f_{\lambda i}(\gamma_i\gamma_j+1) & -\frac g2 f_{\lambda i}(\gamma_i\gamma_j-1)  \\
  \frac g2 f_{\lambda i}(\gamma_i\gamma_j-1)  & \frac g2 f_{\lambda i}(\gamma_i\gamma_j+1)- 2E_i \delta_{ij} \\
\end{array}%
\right)\,.
\end{equation}
Equation (\ref{C_RPA = sinh/sinh}) is more practical since it does not contain an infinite product and we have used it in our calculations below.

Equations (\ref{C_RPA = prod 1/det}) and (\ref{C_RPA = sinh/sinh}) are valid not only for the restricted partition function (\ref{Z_lambda}) with $\lambda = \eta, S_z$, but also for the grand-canonical partition function~\cite{Puddu1991,Lauritzen1993, Rossignoli1998} and for the number-parity-projected partition function.~\cite{Rossignoli1998} In all of these cases, the correct  $Z_\lambda(\Dstaticarg)$ must be used in Eq.~(\ref{F_eta_i}) to define $f_{\lambda i}$. These expressions, however, are not applicable for the canonical projection.

\subsection{Summary}

In summary, we use Eqs.~(\ref{Z^(J)}) and (\ref{chi}) to express the $N$-particle partition function $Z_N$ and spin susceptibility $\chi$ of a system described by the universal Hamiltonian (\ref{universal_hamiltonian}) in terms of the number-parity and $S_z$-projected partition function $Z_{N,\eta,S_z}$ of a system described by the BCS-like pairing Hamiltonian (\ref{BCS_Hamiltonian}) [note that in Eqs.~(\ref{Z^(J)}) and (\ref{chi}) we replace $Z_{S_z}$ by $Z_{N,\eta,S_z}$].  The partition function $Z_{N,\eta,S_z}$ is calculated in the SPA+RPA using
\begin{multline}\label{Z_N}
Z_{N,\eta, S_z}  \approx\int \limits_0^\infty \frac{\beta \,d\,\Dstatic^2}{g} \left(\frac{2\pi}{\beta}\left|\frac{\partial^2 F}{\partial \mu^2}\right|\right)^{-1/2} \\
\times e^{-(\beta/g)\Dstatic^2} \,\, e^{-\beta \mu N}  Z_{\eta, S_z}(\Dstaticarg) \,\,C^{\mathrm{RPA}}_{\eta, S_z}(\Dstaticarg)\,.
\end{multline}
Here $\Dstaticarg$ denotes a static fluctuation of the order parameter, and $\eta$ is the number parity ($\eta= 1$ for even $N$ and $-1$ for odd $N$).
The partition function $Z_{\eta, S_z}(\Dstaticarg)$ and the RPA correction $C^{\mathrm{RPA}}_{\eta, S_z}(\Dstaticarg)$ are given by Eqs.~(\ref{Z(Dstatic)}) and (\ref{C_RPA = prod 1/det}) or  (\ref{C_RPA = sinh/sinh}), respectively. The second partial derivative $\partial^2 F/\partial \mu^2$ is given by Eq.~(\ref{d2Fdmu2}), and the chemical potential $\mu$ for a given static fluctuation $\Dstaticarg$ is determined from Eq.~(\ref{N=-dF/dmu}). The heat capacity $C$ is obtained numerically from the partition function $Z_N(T)$ as a function of temperature.

\begin{figure}[t]
\includegraphics[height=190pt]{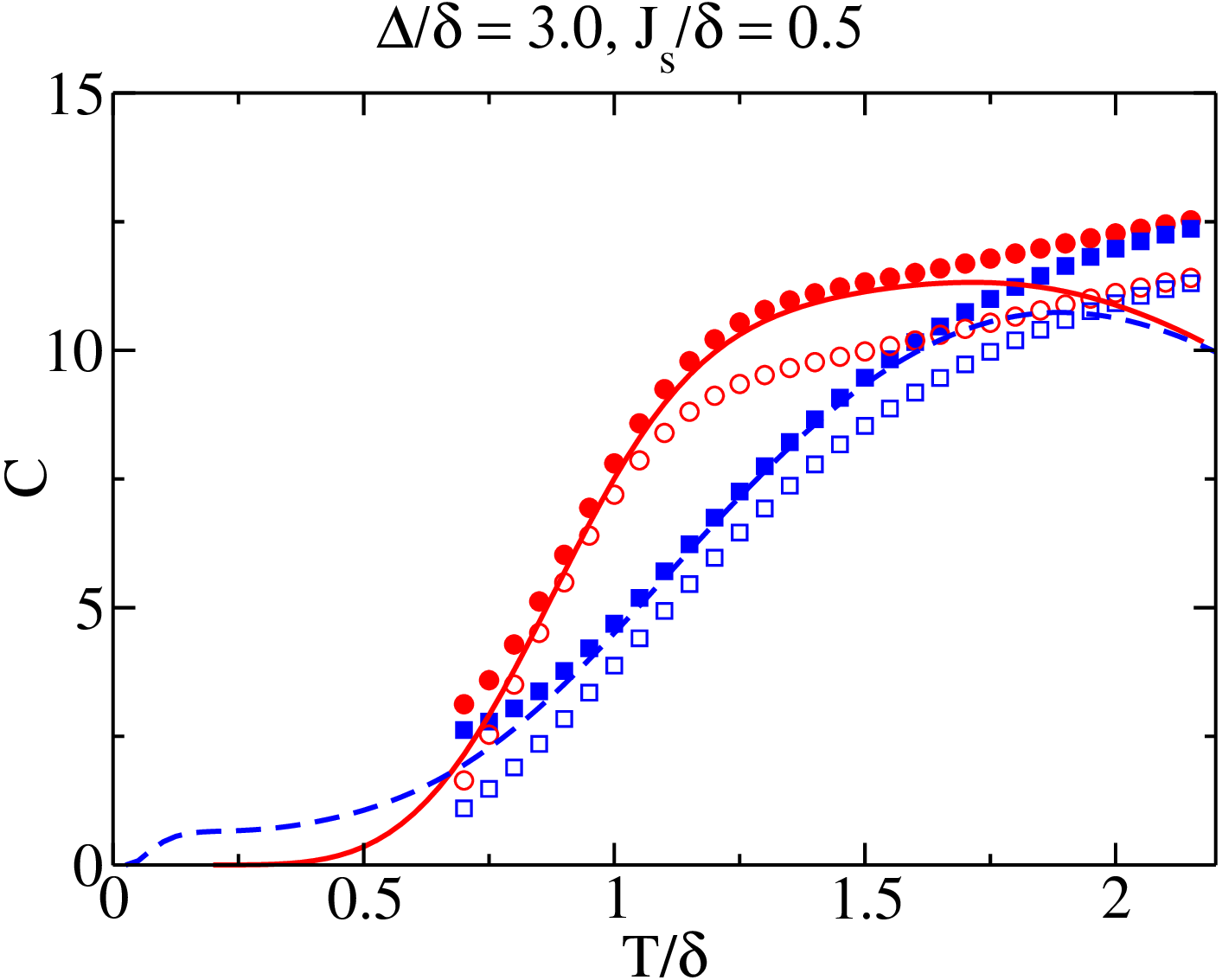}
\caption{Heat capacity $C$ vs $T/\delta$ for  even and odd grains with $\Delta/\delta = 3.0$, $J_s/\delta = 0.5$, and for a specific realization of the GOE single-particle spectrum in (\ref{universal_hamiltonian}). The results calculated in the number-parity projected SPA (open symbols) and SPA+RPA (solid symbols) for even (circles) and odd (squares) grains are compared with exact canonical results obtained by Richardson's solution (solid line for the even grain and dashed line for the odd grain). The results based on Richardson's solution use all eigenvalues below a cutoff of $\sim 30\,\delta$ and are no longer accurate for temperatures above $T \sim 1.5\,\delta$.}\label{hc_Rich}
\end{figure}

\begin{figure}[t]
\includegraphics[height=190pt]{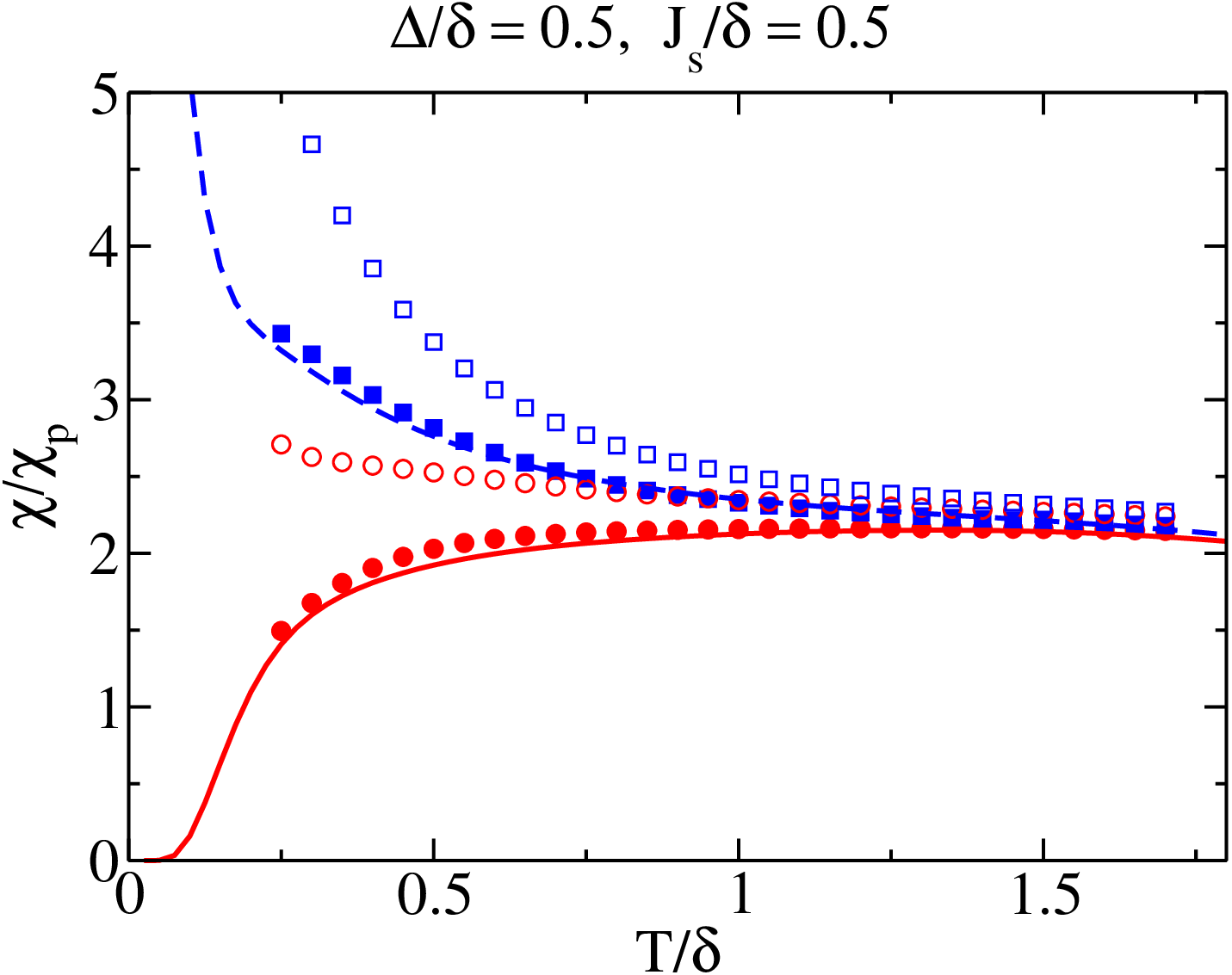}
\caption{As in Fig.~\ref{hc_Rich} but for the spin susceptibility $\chi$ normalized by the Pauli susceptibility $\chi_P = 2\mu^2_B/\delta$ vs $T/\delta$ and for a grain with $\Delta/\delta = 0.5$, $J_s/\delta = 0.5$. Symbols and lines follow the same convention as in Fig.~\ref{hc_Rich}.}\label{ss_Rich}
\end{figure}

\section{Results and Discussion\label{section_results}}

\subsection{Accuracy of the method}

We first discuss the accuracy of the number-parity projected SPA and SPA+RPA methods. To this end, we have used the exact solution of the Hamiltonian (\ref{universal_hamiltonian}), modifying Richardson's solution for  the BCS-like Hamiltonian~\cite{Richardson1963, Richardson1967} to include the exchange interaction.~\cite{Schmidt2007}  The number of many-body eigenstates that contribute to the partition function increases rapidly with temperature, and so does the required computational effort. In practice, we compute only the energy eigenvalues below a cutoff of $\sim 30\,\delta$. These exact calculations are then accurate at sufficiently low temperatures where the contribution of eigenstates with energy above $30\,\delta$ is negligible.

The comparison between the exact and approximate calculations is demonstrated for a given realization of the single-particle spectrum in Figs.~\ref{hc_Rich} and \ref{ss_Rich} for the heat capacity and  spin susceptibility, respectively. We show results for both even and odd number of electrons.

We observe that the number-parity projected SPA+RPA (solid symbols) improves significantly the number-parity projected SPA (open symbols) and provides accurate results for both the heat capacity and spin susceptibility. In particular, the RPA correction is important for the spin susceptibility at larger values (closer to 1) of the exchange coupling $J_s/\delta$.  The example shown on Fig.~\ref{ss_Rich} demonstrates that the number-parity projected SPA results can be qualitatively wrong, whereas the inclusion of the RPA correction factor gives much more accurate results.

\subsection{Heat capacity}

\begin{figure}[t]
\includegraphics[width=240pt]{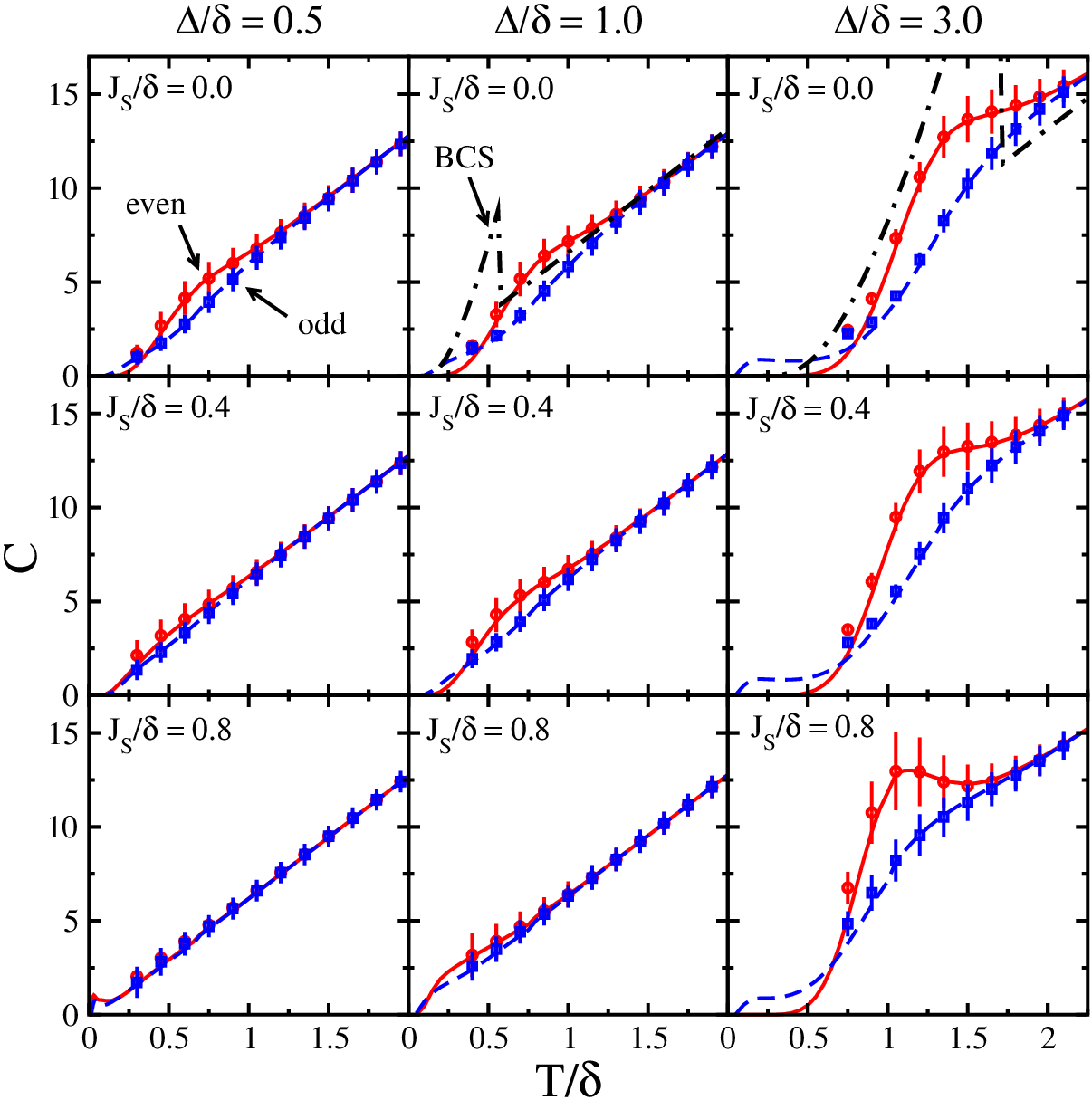}
\caption{The heat capacity $C$  vs temperature $T/\delta$ for an even grain (solid lines, circles) and for an odd grain (dashed lines, squares). Results are shown for grains with
$\Delta/\delta=0.5$ (left column), $\Delta/\delta=1.0$ (middle column), and
$\Delta/\delta=3.0$ (right column) and with $J_s/\delta=0$ (top row), $J_s/\delta=0.4$ (middle row), and $J_s/\delta=0.8$ (bottom row). The
symbols and vertical bars describe, respectively, the average value $\overline{C}$ and standard deviation $\delta C$ of the heat capacity (where an ensemble of single-particle spectra are sampled from the GOE).  The lines correspond to an equally spaced single-particle spectrum in the Hamiltonian (\ref{universal_hamiltonian}) and the dash-dotted lines are the grand-canonical BCS results (where applicable).}\label{hc_mes}
\end{figure}

\begin{figure}[t]
\includegraphics[width=240pt]{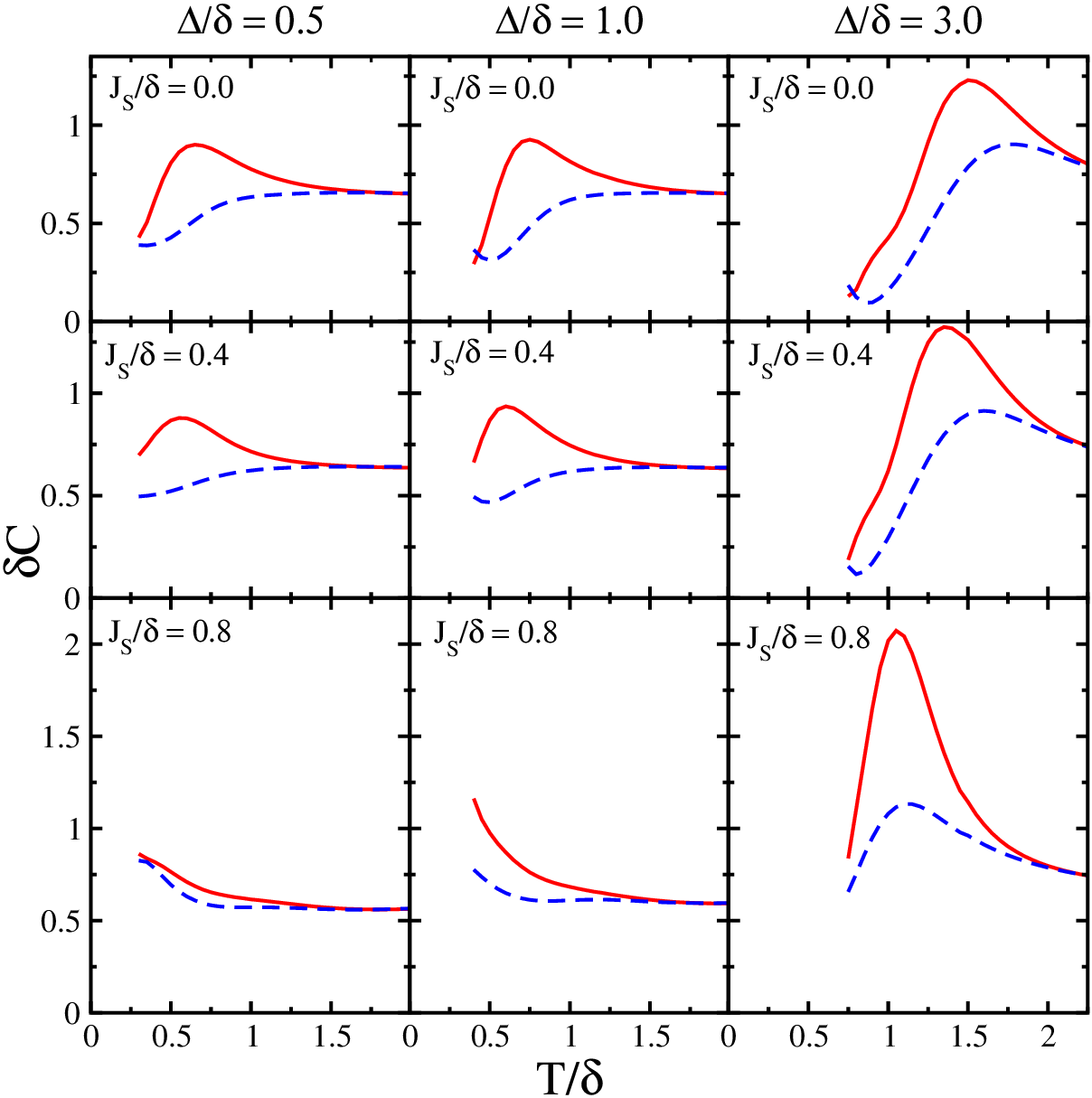}
\caption{The standard deviation $\delta C$ of the heat capacity (shown by vertical bars in Fig.~\ref{hc_mes}) versus temperature $T/\delta$
for an even grain (solid lines) and for an odd grain (dashed lines) with the same values of $\Delta/\delta$ and $J_s/\delta$ as in Fig.~\ref{hc_mes}.
}\label{hc_var}
\end{figure}

There are two major number-parity-dependent signatures of pairing correlations in the heat capacity: the heat capacity for an even particle number is suppressed at low temperatures and enhanced at intermediate temperatures when compared with the heat capacity for an odd particle number (see Richardon's solution results in Fig.~\ref{hc_Rich}). The low-temperature effect is not accessible by the method we are using because the RPA becomes unstable at low temperatures, and we concentrate below on number-parity effects in the intermediate temperature region. It is known that, in the absence of exchange, the characteristic temperature of this region is determined by the scale that is the largest between $\delta$ and $\Delta$.~\cite{Alhassid2007} In the BCS regime (i.e., large $\Delta/\delta$), this effect occurs around the BCS critical temperature, while in the fluctuation-dominated regime $\Delta/\delta \lesssim 1$ it occurs at temperatures higher than the BCS critical temperature. The even-odd effect becomes more prominent when the size of the grain and consequently $\Delta/\delta$  increase.  The heat capacity for even particle number starts displaying a shoulder around $\Delta/\delta \approx 3.0$, which eventually develops into a sharp peak in the bulk limit $\Delta/\delta \gg 1$. Here we investigate how this picture is affected by a non-zero exchange interaction and by mesoscopic fluctuations.

The results for the heat capacity are shown versus $T/\delta$ and both number parities in Fig.~\ref{hc_mes} for $\Delta/\delta = 0.5, 1.0, 3.0$ and $J_s/\delta = 0, 0.4, 0.8$.  The symbols and vertical bars are average values $\overline{C}$ and standard deviations $\delta C$, respectively, calculated from an ensemble of 1000 random matrices describing the one-body part of the Hamiltonian
(\ref{universal_hamiltonian}). The lines (solid for even and dashed for odd number of electrons)  are obtained for the equally spaced single-particle spectrum in the Hamiltonian (\ref{universal_hamiltonian}) with level spacing $\delta$. Figure~\ref{hc_var} shows the standard deviation $\delta C$ versus $T/\delta$ for the same cases as in Fig.~\ref{hc_mes}.

We observe that the exchange interaction can suppress the odd-even effects in the heat capacity and shift them to lower temperatures. This is particularly evident if $\Delta/\delta \lesssim 1$. For $\Delta/\delta = 3.0$, a higher value of the exchange coupling constant is required to make a visible change. Even for $J_s/\delta = 0.8$ (which is close to the Stoner instability threshold), the number-parity effect is shifted to lower temperatures  slightly. Only the right side of the even number-parity shoulder is suppressed, whereas the left side is not. As a result, the shoulder transforms into a peak.

This behavior is consistent with the ground-state phase diagram~\cite{Schmidt2007} of the grain, according to which the ground state for an even particle number is fully paired for small  $J_s/\delta$ and the value of $J_s/\delta$ required to polarize the grain increases with  $\Delta/\delta$. For $\Delta/\delta = 3.0$, this value of $J_s/\delta$ is close to the Stoner instability  threshold. For smaller values of $J_s/\delta$, the excited states with non-zero spin become important only at sufficiently high temperatures. At lower temperatures, the dominant contribution to the heat capacity comes from the zero spin levels whose energy is independent of $J_s/\delta$. This is consistent with  the weak dependence on $J_s/\delta$ of the left side of the even-case shoulder for $\Delta/\delta = 3.0$. At temperatures that correspond to the right side of the even-case shoulder, non-zero spin configurations are more important and lead to visible change in the heat capacity. For $\Delta/\delta \lesssim 1$, the excitation energies of the states with non-zero spin are lower and the heat capacity is more sensitive to exchange correlations at lower temperatures.

In an ultra-small grain, the mesoscopic fluctuations of observables can be significant. For example, if an odd-even signature of pairing correlations is studied by carrying out measurements in many samples with unknown number parity and then   determining the distribution of the observable, such a number-parity effect may be washed out when the fluctuations are large. As can be seen from our results, this can happen if $\Delta/\delta$ is sufficiently small or $J_s/\delta$ is sufficiently large. In the absence of exchange, this occurs at $\Delta/\delta \lesssim 0.5$, and, for $J_s/\delta \sim 0.5$, number-parity effects already disappear below $\Delta/\delta \sim 1$. However, for $\Delta/\delta = 3.0$, the odd-even effect is not suppressed by the fluctuations even at relatively large $J_s/\delta$.

As can be seen in Fig.~\ref{hc_var} the mesoscopic fluctuations of the heat capacity at high temperatures are only weakly dependent on the pairing gap, exchange coupling constant and the number parity. At intermediate temperatures, when the heat capacity is enhanced for an even particle number, the fluctuations in the even case are stronger than in the odd case and are characterized by a peak. The position and height of this peak are almost independent of $\Delta/\delta$ for $\Delta/\delta \lesssim 1$,  but increase with $\Delta/\delta$ for $\Delta/\delta >1$. In the presence of exchange correlations, the peak is shifted to lower temperatures.  This is consistent with a similar shift of the odd-even signature in the heat capacity. In addition the size of the peak increases with $J_s$ for $\Delta/\delta > 1$.

\subsection{Spin susceptibility}

\begin{figure}[t]
\includegraphics[width=240pt]{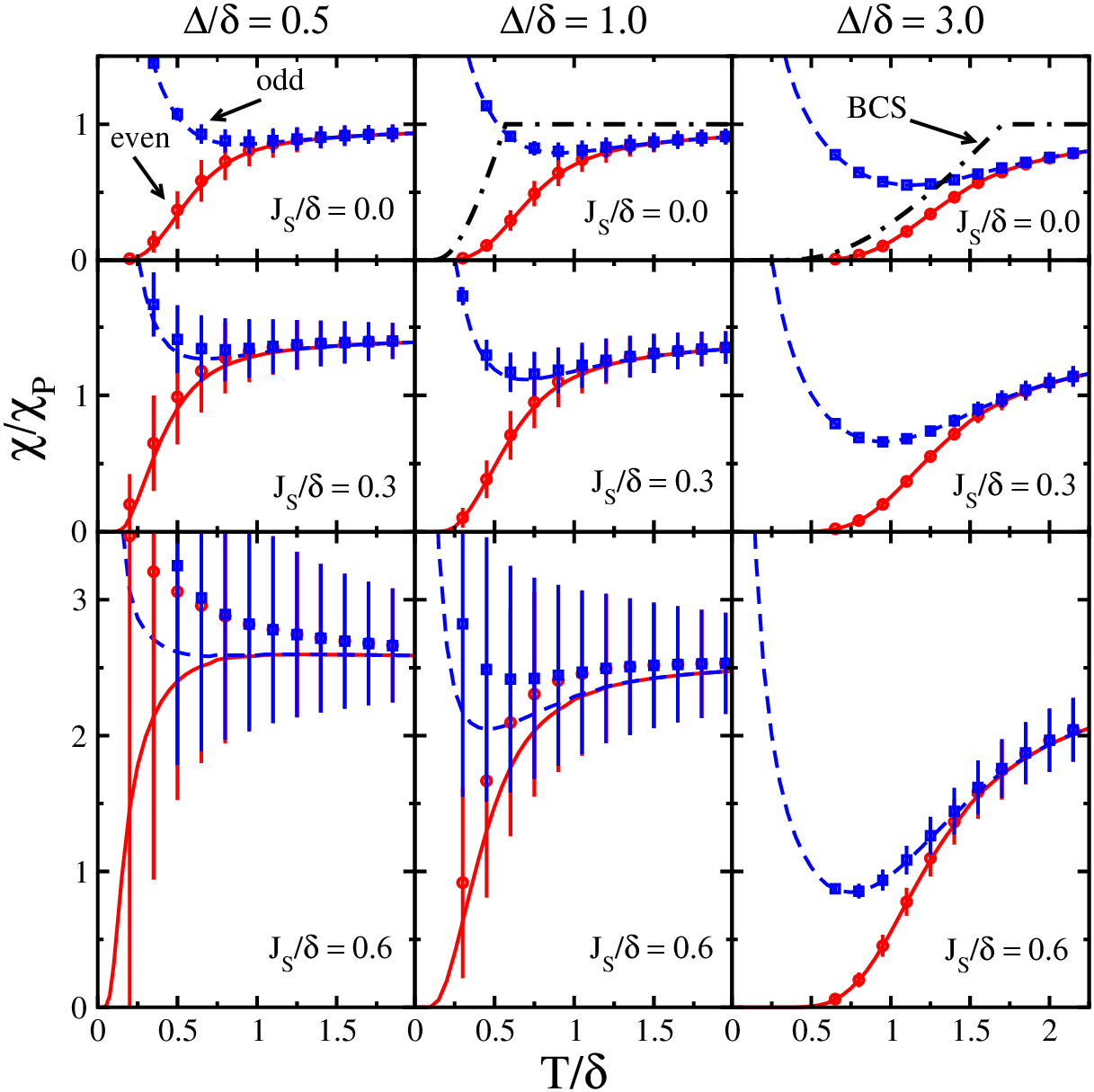}
\caption{The spin susceptibility $\chi$ in the units of the Pauli susceptibility  $\chi_P = 2\mu^2_B/\delta$ vs
temperature $T/\delta$ for even and odd grains with
$\Delta/\delta=0.5$ (left column), $\Delta/\delta=1$ (middle column) and
$\Delta/\delta=3.0$ (right column) and with $J_s/\delta=0$ (top row), $J_s/\delta=0.3$ (middle row) and $J_s/\delta=0.6$ (bottom row). Symbols and lines follow the same convention as in Fig.~\ref{hc_mes} but for the spin susceptibility.}\label{ss_mes}
\end{figure}

\begin{figure}[t]
\includegraphics[width=240pt]{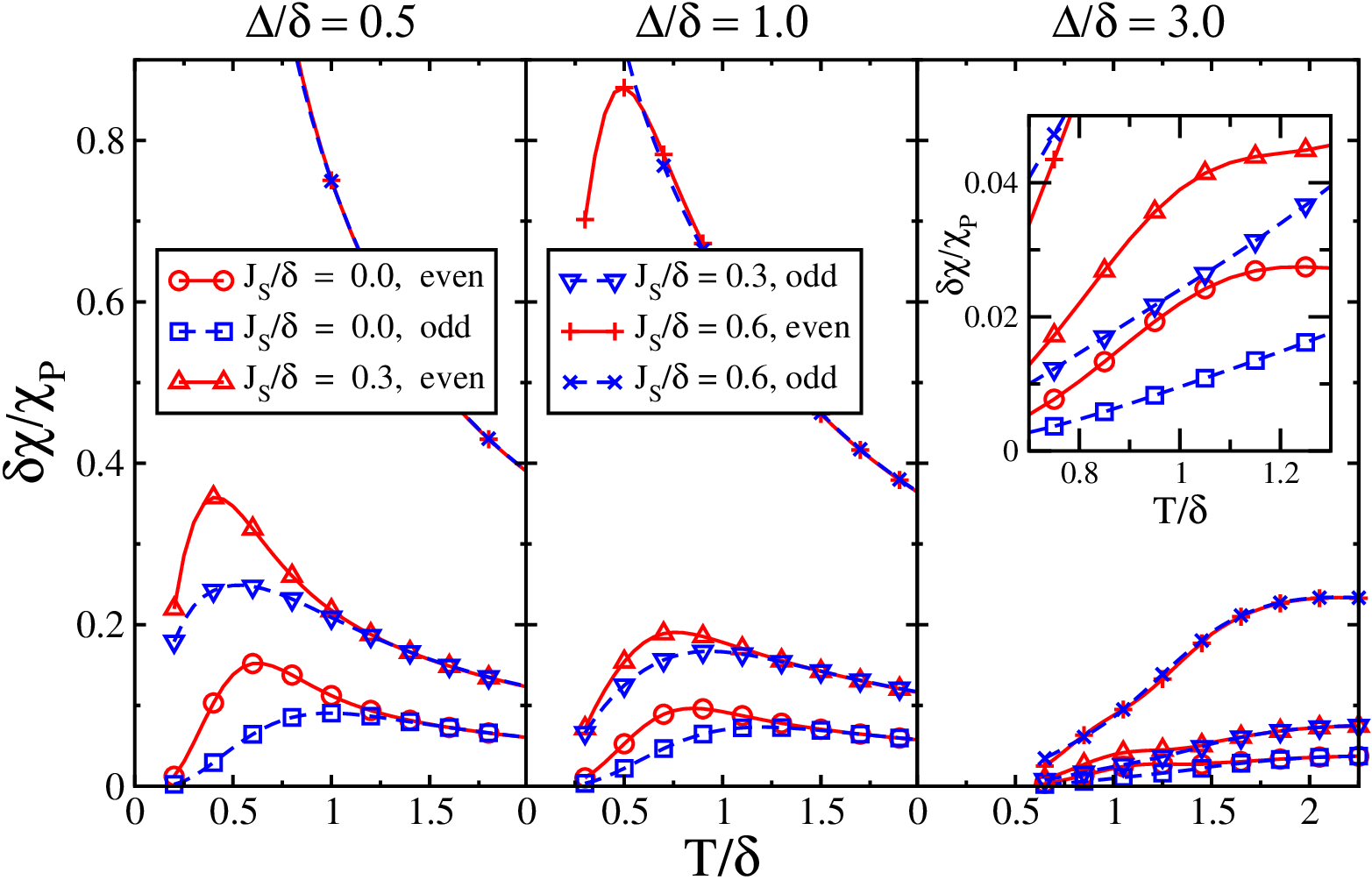}
\caption{The standard deviation $\delta \chi/\chi_P$ of the spin susceptibility
(shown by vertical bars in Fig.~\ref{ss_mes}) vs temperature $T/\delta$
for an even grain (solid lines) and for an odd grain (dashed lines) with the same values of $\Delta/\delta$ and $J_s/\delta$ as in Fig.~\ref{ss_mes}.}\label{ss_var}
\end{figure}

\begin{figure}[t]
\includegraphics[width=240pt]{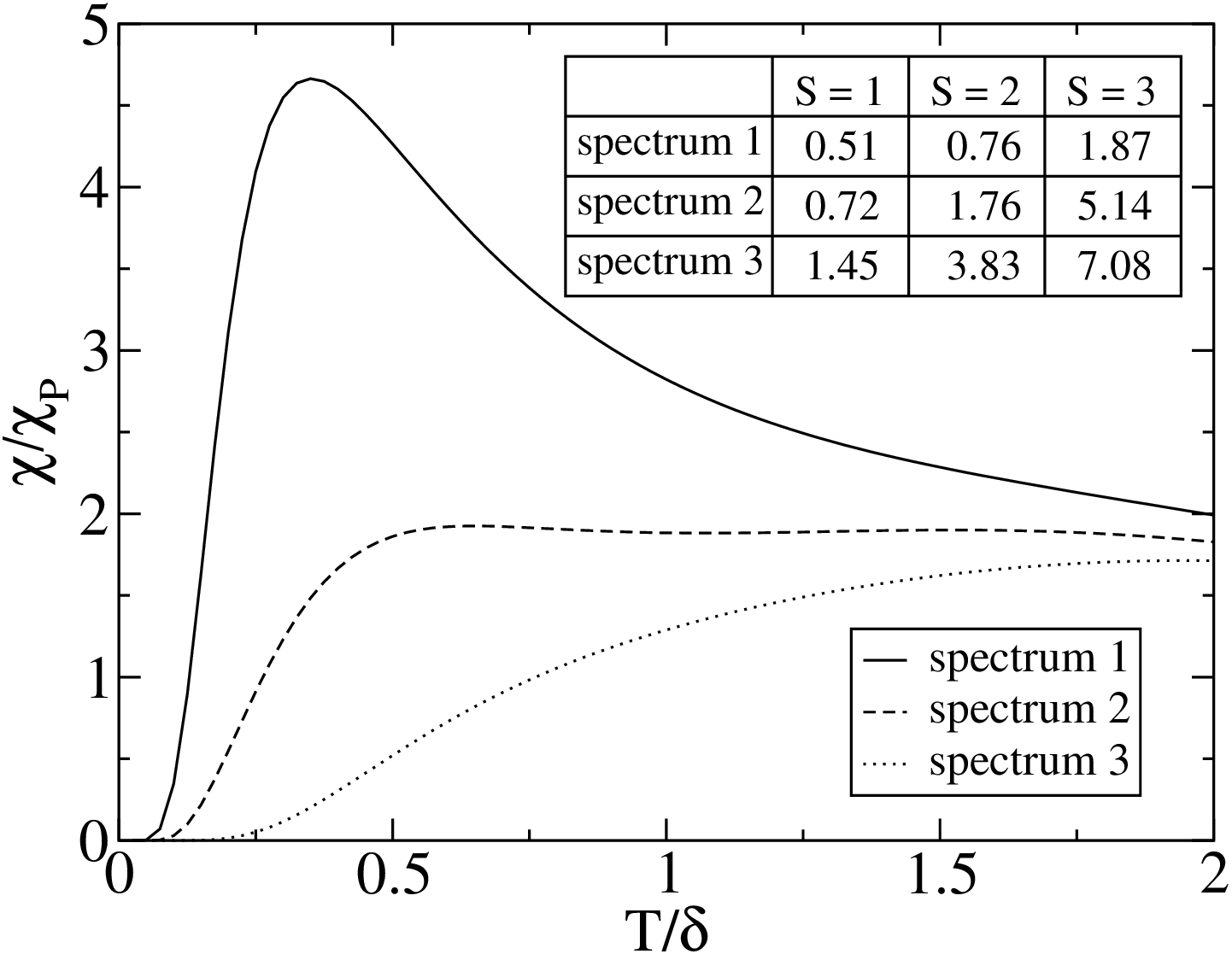}
\caption{The spin susceptibility $\chi/\chi_P$ for an even grain with $\Delta/\delta = 0.5$ and $J_s/\delta = 0.5$ for three different RMT realizations of the single-particle spectrum in (\ref{universal_hamiltonian}). The table shows (for each of the three realizations) the lowest excitation energy $E_S/\delta$ for a given spin $S$.}\label{ss_rms}
\end{figure}

It is known that pairing correlations (in the absence of exchange) suppress the spin susceptibility for both number parities. For an odd particle number, this suppression together with the low-temperature Curie-like divergence ($\sim 1/T$) leads to a re-entrant behavior. For an even particle number, the spin susceptibility increases monotonically with temperature.  Exchange correlations and mesoscopic fluctuations may affect this behavior.

We express the spin susceptibility $\chi$ in the units of the Pauli susceptibility  $\chi_P = 2\mu^2_B/\delta$, where $\mu_B$ is the Bohr magneton. For given values of  $\Delta/\delta$ and $J_s/\delta$, the ratio $\chi/\chi_P$ is expected to be a universal function of $T/\delta$. In the high-temperature limit, $\chi/\chi_P$ does not depend on $\Delta/\delta$. For an equally-spaced single-particle spectrum and in the absence of exchange, it approaches $1$ in that limit.

The results for the spin susceptibility versus $T/\delta$ are shown (for both number parities) in Fig.~\ref{ss_mes} for $\Delta/\delta = 0.5,\, 1.0,\, 3.0$, and $J_s/\delta= 0,\, 0.3
,\, 0.6$. Symbols and lines follow the same convention as in Fig.~\ref{hc_mes}.  The standard deviation $\delta\chi/\chi_P$ is shown in Fig.~\ref{ss_var} versus $T/\delta$.

The most visible effect of the exchange interaction is the enhancement of $\chi/\chi_P$ as higher spin states are shifted down in energy. Exchange correlations can also shift the odd-even effects to lower temperatures and even  eliminate the odd-case re-entrant behavior for $\Delta/\delta \lesssim 1$. However, at larger values of $\Delta/\delta$, exchange enhances the re-entrant behavior. This effect is similar to what we observed for the even-case heat capacity where a shoulder changes into a peak for larger values of $\Delta/\delta$ (see the right column of Fig.~\ref{hc_mes}).

The mesoscopic fluctuations of $\chi/\chi_P$ increase with decreasing $\Delta/\delta$ or increasing $J_s/\delta$. In the fluctuation-dominated regime $\Delta/\delta \lesssim 1$, they can become especially strong at larger values of $J_s/\delta$ and can hinder the observation of odd-even effects. When compared to the heat capacity results, higher values of $J_s/\delta$ or smaller values of $\Delta/\delta$ are required to hinder the odd-even effects.

The large mesoscopic fluctuations of the spin susceptibility for $\Delta/\delta \lesssim 1$ and large values of $J_s/\delta$ may be attributed to the large dispersion of the magnetization of the system. This is confirmed by studying spectra of individual samples in that regime using Richardon's method. Examples are shown in  Fig.~\ref{ss_rms}. Large spin susceptibility values are obtained in samples in which the excitation energies of states whose spin is different from the ground-state spin are particularly low. The probability to have such a sample is enhanced when pairing correlations are weaker and/or exchange correlations are larger.

At relatively large values of $\Delta/\delta$, the fluctuations increase monotonically with temperature (see right column of Fig.~\ref{ss_var}). However, for $\Delta/\delta \lesssim 1$, the fluctuations in the spin susceptibility have a maximum at smaller values of  $J_s/\delta$.
It is not clear if this is the case at large values $J_s/\delta$ because we cannot access very low temperatures by our method. However, we expect the fluctuations to remain strong in the limit $T \rightarrow 0$ in the odd case because of the  Curie-like behavior $\sim S_0/T$, where $S_0$ is fluctuating ground-state spin.

We note that the equally spaced single-particle spectrum does not always describe the average behavior of the spin susceptibility. In the regime where the mesoscopic fluctuations are strong, the spin susceptibility calculated for the equally-spaced spectrum (solid lines in Fig.~\ref{ss_mes}) is smaller than the spin susceptibility obtained by averaging over all samples (circles) and may be qualitatively different (see, e.g., the case $\Delta/\delta = 0.5$ and $J_s/\delta$ = 0.6 in Fig.~\ref{ss_mes}).

\section{Conclusions \label{section_conclusions}}

In conclusion, we have studied the thermodynamic properties of ultra-small chaotic metallic grains with a large dimensionless Thouless conductance in the presence of both superconducting and ferromagnetic correlations. We have used the so-called universal Hamiltonian (\ref{universal_hamiltonian}) as our model, in which the one-body part is sample-specific and modeled by RMT, while the dominating interaction terms are universal. Sample-to-sample fluctuations of the interaction are suppressed and ignored in the limit of large dimensionless  Thouless conductance. The exchange interaction has been treated exactly by means of a spin-projection method, while the pairing interaction has been treated in a path-integral approach in which all static fluctuations of the pairing gap and small-amplitude time-dependent fluctuations around each static value of the gap are included (SPA+RPA method). Particle-number projection is approximated in the saddle-point approximation while number-parity effects are preserved using an exact number-parity projection. The method is efficient and very accurate when compared to exact canonical calculations. However, it cannot be used at very low temperatures, when the RPA correction becomes unstable. This limitation can potentially be overcome using the method developed in Ref.~\onlinecite{Ribeiro2012}.

We have found that the exchange interaction shifts the number-parity-dependent signatures of pairing correlations (such as the enhancement of heat capacity in the even grain and the re-entrant behavior of the spin susceptibility in the odd grain) to lower temperatures. In the fluctuation-dominated regime $\Delta/\delta \lesssim 1$, these signatures are suppressed by exchange correlations. However, at sufficiently large values of $\Delta/\delta$,  exchange correlations have the opposite effect, i.e., the heat capacity of the even grain develops a peak, and the re-entrant behavior of the spin susceptibility in the odd grain is enhanced.

Mesoscopic fluctuations of thermodynamic observables can further hinder the odd-even effects for sufficiently small $\Delta/\delta$ and large $J_s/\delta$.
The mesoscopic fluctuations of the spin susceptibility are especially large in the fluctuation-dominated regime $\Delta/\delta \lesssim 1$ for values of $J_s/\delta$ above $\sim 0.5$ because of the large dispersion of the magnetization.

It would be interesting to extend our work to the study of granular metals,~\cite{Beloborodov2007} i.e., arrays of metallic nanoparticles that are coupled via tunnel junctions. For weakly coupled grains and  when the charging energy $E_C$ satisfies $E_C \gg \Delta$, the majority of grains are in the Coulomb-blockade regime $k T \ll E_C$ with suppressed inter-grain tunneling. These Coulomb-blockaded grains provide the dominant contribution to the thermodynamic properties of the granular metal at low temperatures. Therefore, the values of thermodynamic observables of a granular metal (per grain) at $k T\ll E_C$ can be effectively calculated by averaging the observables of individual grains over different random-matrix realizations and the number parity of electrons. We note that number-parity effects must still be taken into account since they lead to effects that could be missed in grand-canonical calculations.  An example is the Curie-like divergence in the average spin susceptibility.

\begin{acknowledgments}

We thank Sebastian Schmidt for useful discussions. This work was supported in part by the U.S. DOE grant No. DE-FG02-91ER40608, and by the facilities and staff of the Yale University Faculty of Arts and Sciences High Performance Computing Center.

\end{acknowledgments}

\appendix

\section{Calculation of the RPA correction factor} \label{Appendix_RPA_correction}

In this appendix we calculate the RPA correction factor  (\ref{C_RPA_definition}) with $\lambda$ denoting  simultaneous projections on the number parity $\eta$ and spin component $S_z$. The corresponding projection operators, given by Eqs.~(\ref{P_eta_definition}) and (\ref{P_S_z}), respectively, have the same form in both single-particle and $\Dstaticarg$-dependent quasiparticle representations. This indicates that they commute with the quasiparticle occupation number operator $a^\dagger_{i\sigma}a_{i\sigma}$ for each $i$ and $\sigma$. Therefore, if $\Vtimedepint(\tau)$ is written in the quasiparticle representation according to Eq.~(\ref{V_sigma_quasiparticle_basis}) with $\tau$-dependent operators, non-zero contributions to the correlator of $\Vtimedepint(\tau)$ in Eq.~(\ref{C_RPA_definition}) 
are possible only from  the product of  $a^\dagger_{i\uparrow}a^\dagger_{i\downarrow}$ and  $a_{i\downarrow}a_{i\uparrow}$ and from the product of two terms of the form $1-a^\dagger_{i\downarrow}a_{i\downarrow} - a^\dagger_{i\uparrow}a_{i\uparrow}$ taken from both $\Vtimedepint(\tau)$ and $\Vtimedepint(\tau')$. In the latter case, the two terms are $\tau$-independent and identical. Thus time-ordering can be omitted, and integration over $\tau$ vanishes because of the $e^{i\omega_r\tau}$ factor.  Consequently, we obtain
\begin{multline}\label{B-integral}
\frac 12 \int\limits_0^\beta \int\limits_0^\beta d\tau d\tau' \left\langle {\cal T}\,\Vtimedepint(\tau) \Vtimedepint(\tau') \right\rangle_{\lambda,\Dstaticarg} \\
= \frac{1}{4} \sum_{i}\sum_{r,r'\neq 0} \Lambda_{ir}\Lambda^*_{ir'} \int\limits_0^\beta \int\limits_0^\beta d\tau d\tau' e^{i\omega_r\tau-i\omega_{r'}\tau'}B_i(\tau,\tau')\,,
\end{multline}
where
\begin{equation}
  B_{i}(\tau,\tau') = \left\langle {\cal T}\,\,
a_{i\uparrow}^\dagger(\tau)a_{i\downarrow}^\dagger(\tau)a_{i\downarrow}
(\tau')a_{i\uparrow}(\tau')\right\rangle_{\lambda,\Dstaticarg}\,.
\end{equation}

Wick's theorem cannot be applied directly to the correlator $B_i(\tau,\tau')$ because of the projections. To proceed, we use the following two identities
\begin{equation}\label{appendix_identity_1}
\left\langle a_{i\downarrow}a_{i\uparrow} a^\dagger_{i\uparrow}a^\dagger_{i\downarrow}\right\rangle_{\lambda,\Dstaticarg} = e^{2\beta E_i} \left\langle a^\dagger_{i\uparrow} a^\dagger_{i\downarrow} a_{i\downarrow}a_{i\uparrow}\right\rangle_{\lambda,\Dstaticarg}
\end{equation}
and
\begin{multline}\label{appendix_identity_2}
(1-e^{2\beta E_i})\left\langle a^\dagger_{i\uparrow} a^\dagger_{i\downarrow}a_{i\downarrow} a_{i\uparrow}\right\rangle_{\lambda,\Dstaticarg} \\
= -1 + \left\langle a^\dagger_{i\downarrow}a_{i\downarrow} \right\rangle_{\lambda,\Dstaticarg} + \left\langle a^\dagger_{i\uparrow}a_{i\uparrow} \right\rangle_{\lambda,\Dstaticarg}\,.
\end{multline}

The second identity follows directly from the first one and the anti-commutation relations. To derive the first identity, we write the projected average (\ref{average_lambda}) of an observable $\hat A$  at a given static field in the form
\begin{equation}\label{lambda-projection}
\left\langle \hat{A}\right\rangle_{\lambda,\Dstaticarg} = \sum_{\phi_\lambda} \widetilde{C}_{\phi_\lambda} \left\langle\hat{A}\right\rangle_{\phi_\lambda}\;,
\end{equation}
 where $\widetilde{C}_{\phi_\lambda}$ are certain coefficients and
\begin{equation}\label{Fourier-A}
\left\langle\hat{A}\right\rangle_{\phi_\lambda} = \frac{\Tr\left(\USPA e^{i\phi_\lambda\hat{S}_z} \hat{A}\right)}{\Tr\left(\USPA e^{i\phi_\lambda\hat{S}_z} \right)} \;.
\end{equation}
The sum in Eq.~(\ref{lambda-projection}) is over quadrature points $\phi_\lambda$ suitable for the projection $\lambda$.  In deriving (\ref{lambda-projection}), we have used expressions (\ref{P_eta_definition}) and (\ref{P_S_z}) for the projection operators and replaced $e^{i\pi\hat{N}}$ by $e^{2i\pi\hat{S}_z}$ ($N$ and $2S_z$ have the same parity).
The expectation value in Eq.~(\ref{Fourier-A}) is grand canonical with respect to the one-body Hamiltonian $\Hstatic-i\phi_\lambda \hat{S}_z/\beta$ ($\Hstatic$ commutes with $\hat{S}_z$), and therefore can be calculated using Wick's theorem. We find
\begin{multline}\label{appendix_identity_1a}
\left\langle a_{i\downarrow}a_{i\uparrow} a^\dagger_{i\uparrow}a^\dagger_{i\downarrow}\right\rangle_{\phi_\lambda} = (1-n_{i\uparrow}^{\phi_\lambda})(1-n_{i\downarrow}^{\phi_\lambda}) \\
=e^{2\beta E_i } n_{i\uparrow}^{\phi_\lambda} n_{i\downarrow}^{\phi_\lambda} = e^{2\beta E_i}\left\langle  a^\dagger_{i\uparrow}  a^\dagger_{i\downarrow}a_{i\downarrow}a_{i\uparrow}\right\rangle_{\phi_\lambda}\,.
\end{multline}
where
\begin{equation}
n_{i\sigma}^{\phi_\lambda} = \left\langle a^\dagger_{i\sigma}a_{i\sigma}\right\rangle_{\phi_\lambda} = \frac{1}{e^{\beta E_i \mp i\phi_\lambda/2}+1} \;.
\end{equation}
Using Eqs.~(\ref{lambda-projection}) and (\ref{appendix_identity_1a}), we obtain the relation (\ref{appendix_identity_1}).

We can now evaluate the integrals on the right-hand side of Eq.~(\ref{B-integral}) with the help of Eq.~(\ref{a(tau)=a*e}) and the identities  (\ref{appendix_identity_1}) and (\ref{appendix_identity_2}) to find
\begin{widetext}
\begin{multline}
\int\limits_0^\beta \int\limits_0^\beta d\tau d\tau' e^{i\omega_r\tau-i\omega_{r'}\tau'}B_i(\tau,\tau')
=\left\langle a_{i\uparrow}^\dagger  a_{i\downarrow}^\dagger a_{i\downarrow} a_{i\uparrow} \right\rangle_{\lambda\Dstaticarg}
  \int\limits_0^\beta \int\limits_0^\beta d\tau d\tau' e^{i\omega_r\tau-i\omega_{r'}\tau'} e^{2E_i(\tau-\tau')}\left[\theta(\tau-\tau') +\theta(\tau'-\tau)e^{2\beta E_i}\right]   \\
= \left\langle a_{i\uparrow}^\dagger  a_{i\downarrow}^\dagger a_{i\downarrow} a_{i\uparrow} \right\rangle_{\lambda\Dstaticarg} \int\limits_0^\beta d\tau e^{(i\omega_r+2E_i)\tau}  \left[-\frac{e^{-(i\omega_{r'}+2E_i)\tau} - 1}{i\omega_{r'}+2E_i} - \frac{1-e^{2\beta E_i} e^{-(i\omega_{r'}+2E_i)\tau}}{i\omega_{r'}+2E_i} \right]
= \frac{\beta \delta_{rr'}}{i\omega_r+2E_i} f_{\lambda i}\,,
\end{multline}
\end{widetext}
where
\begin{equation}\label{F_lambda_appendix}
f_{\lambda i} = 1 - \left\langle\left( a^\dagger_{i\downarrow}a_{i\downarrow} + a^\dagger_{i\uparrow}a_{i\uparrow} \right)\right\rangle_{\lambda\Dstaticarg} = \frac{1}{\beta} \frac{\partial \ln Z_{\lambda}(\Dstaticarg)}{\partial E_i}\,.
\end{equation}

Denoting by $\sigma_1(\tau)$ and $\sigma_2(\tau)$ the real and imaginary parts of $\Delta(\tau)e^{-i\theta}$, respectively, we change variables
\begin{equation}
\Delta_{r}=e^{i\theta}(\sigma_{1r}+i\sigma_{2r})\,,
\end{equation}
where $\sigma_{kr}$ is the Fourier transform of $\sigma_{k}(\tau)$ ($\sigma_{k,-r} = \sigma^*_{kr}$). The quantity $\Lambda_{ir}$ in Eq.~(\ref{Lambda_ir}) can then be written as
\begin{equation}
\Lambda_{ir} = 2e^{i\theta} \left(\gamma_i \sigma_{1r} + i\sigma_{2r}\right)\,
\end{equation}
and the integration measure as
\begin{equation}
\mathcal{D}'[\Delta_r,\Delta^*_r] = \mathcal{D}'[\sigma_{1r},\sigma_{2r}] = \prod_{r>0} \frac{\beta^2 d\sigma_{1r}d\sigma^*_{1r}d\sigma_{2r}d\sigma^*_{2r}}{\pi^2g^2}\,.
\end{equation}
Consequently, the RPA correction factor is given by
\begin{widetext}
\begin{multline}\label{C_RPA = det^-1}
C_\lambda^{\text{RPA}}(\Dstaticarg)
= \int \mathcal{D}'[\sigma_{1r},\sigma_{2r}] \exp\left[-(2\beta/g) \sum_{r>
0}\left(\left(1-g\sum_i\frac{2E_i\gamma_i^2{f}_{\lambda
i}}{4E_i^2+\omega_r^2}\right)|\sigma_{1r}|^2+\left(1-g\sum_i\frac{2E_if_{\lambda
i}}{4E_i^2+\omega_r^2}\right)|\sigma_{2r}|^2 + \right.\right.
\\
+ \left.\left. g\sum_i\frac{\omega_r\gamma_i f_{\lambda
i}}{4E_i^2+\omega^2_r}(\sigma_{1r}\sigma^*_{2r}-\sigma_{1r}^*\sigma_{2r})\right)\right]
= \prod_{r>0} \left[\det A(\omega_r)
\right]^{-1}\,,
\end{multline}
\end{widetext}
where the matrix $A(\omega_r)$ is defined in (\ref{A-matrix}). Note that in general Eq.~(\ref{C_RPA = prod 1/det}) is valid as long as Eq.~(\ref{appendix_identity_1}) holds or the  projection operator in $Z_\lambda$ conserves the quasiparticle occupation numbers.

\section{Relation of the RPA correction factor to the RPA matrix}\label{Appendix_RPA_matrix}

As a function of $\omega$, $\det A(\omega)$ in (\ref{C_RPA = det^-1}) has poles at $\omega = \pm 2iE_i$, which could be either first or second order. A  second-order pole at $\pm 2iE_i$ can only arise from products of two terms that contribute to the matrix elements of $A(\omega)$ and have $(4E_i^2+\omega_r^2)$ in their denominators. However, when the sum of these products is written as a ratio of polynomials, a partial cancellation between denominator and numerator results in first-order poles.

Since all the poles of $\det A(\omega)$ are first order, it can be written as
\begin{equation}
\det A(\omega) = \frac{P(\omega)}{\prod_i (\omega^2 + 4E_i^2)}\,,
\end{equation}
where $P(\omega)$ is a polynomial of degree $2N_{\text{sp}}$  ($N_{\text{sp}}$ us the number
of single-particle orbitals). The roots of $P(\omega)$ and $\det A(\omega)$ come in pairs $\pm i\Omega_i$ ($\det A(\omega)$ is a function of $\omega^2$), and the leading coefficient of $P(\omega)$ is equal to one. Thus $P(\omega) = \prod_i (\omega^2 + \Omega_i^2)$, and
\begin{equation}\label{det-A}
\det A(\omega_r) = \prod_i  \frac{\omega_r^2 + \Omega_i^2}{\omega_r^2 + 4E_i^2}\,.
\end{equation}
Using (\ref{C_RPA = det^-1}) and the infinite product representation $\sinh x = x \prod_{r>0} (1+x^2/\pi^2r^2)$, we obtain Eq.~(\ref{C_RPA = sinh/sinh}) for the RPA correction factor.

Next, we show that  $\pm \Omega_i$ are the eigenvalues of the $2N_{\text{sp}}\times 2N_{\text{sp}}$ RPA matrix (\ref{RPA_matrix}). Indeed, considering one of the eigenvalues $\Omega$ of this matrix
and denoting its corresponding eigenvector by $\left( \chi_{1i} \,\,\chi_{2i} \right)^T $, we have
\begin{equation}
\left\{
\begin{aligned}
\chi_{1i} = \frac g2 \frac{f_{\lambda i}}{2E_i -
\Omega}\sum_j\left((\gamma_i\gamma_j+1)\chi_{1j}+(\gamma_i\gamma_j-1)\chi_{2j}\right)\,,\\
\chi_{2i} = \frac g2 \frac{f_{\lambda i}}{2E_i +
\Omega}\sum_j\left((\gamma_i\gamma_j-1)\chi_{1j}+(\gamma_i\gamma_j+1)\chi_{2j}\right)\,.\\
\end{aligned}\right.
\end{equation}
Defining
\begin{equation}
\left\{
\begin{aligned}
\eta_+ &= \sum_j (\chi_{1j}+\chi_{2j})\gamma_j \,,\\
\eta_- &= \sum_j(\chi_{1j}-\chi_{2j})\,,
\end{aligned}\right.
\end{equation}
we obtain
\begin{widetext}
\begin{equation}\label{B-matrix}
\left\{
\begin{aligned}
\eta_+ = \frac g2 \sum_i\left[\frac{f_{\lambda
i}\gamma_i}{2E_i-\Omega}(\gamma_i \eta_+ + \eta_-) +
\frac{f_{\lambda i}\gamma_i}{2E_i+\Omega}(\gamma_i \eta_+ \right.
\left.- \eta_-)\right]
= g\sum_i \frac{2E_i\gamma_i^2f_{\lambda
i}}{4E_i^2-\Omega^2}\,\eta_+ +
g\sum_i \frac{\Omega\gamma_if_{\lambda i}}{4E_i^2-\Omega^2}\,\eta_- \,,\\
\eta_- = \frac g2 \sum_i\left[\frac{f_{\lambda
i}}{2E_i-\Omega}(\gamma_i \eta_+ + \eta_-) -
\frac{f_{\lambda i}}{2E_i+\Omega}(\gamma_i \eta_+  -
\eta_-)\right]
= g\sum_i \frac{\Omega\gamma_if_{\lambda
i}}{4E_i^2-\Omega^2}\,\eta_+ + g\sum_i \frac{2E_if_{\lambda i}}{4E_i^2-\Omega^2}\,\eta_- \,.\\
\end{aligned}\right.
\end{equation}
\end{widetext}
Equation (\ref{B-matrix}) can be rewritten in the form
\begin{equation} \label{determinant_zero}
B(\Omega) \left(
\begin{array}{l}
\eta_+ \\
\eta_-
\end{array} \right) =0\,,
\end{equation}
where $B(\Omega)$ is a $2\times 2$ matrix satisfying $\det B(\Omega) = \det A(i\Omega)$. In general, $\left(\begin{array}{l}\eta_+ \\ \eta_-\end{array} \right) \neq 0$ and Eq.~(\ref{determinant_zero}) implies $\det B(\Omega) = 0$. Therefore, $\omega=i\Omega$ is a root of $\det A(\omega)$. Since the eigenvalues of the RPA matrix (\ref{RPA_matrix}) come in pairs $\pm \Omega$ and their number is equal to the number of roots of $\det A(\omega)$, we conclude that all of the $\Omega_i$ in Eq.~(\ref{C_RPA = sinh/sinh}) are eigenvalues of the matrix (\ref{RPA_matrix}) and vice versa.

The approximation used to calculate the correction factor (\ref{C_RPA = prod 1/det}) breaks down when at least one of the matrices $A(\omega_r)$ has an eigenvalue with negative real part and the corresponding Gaussian integral diverges. As discussed in Sec.~\ref{RPA_correction}, the necessary and sufficient conditions for the SPA+RPA to be applicable are that all matrices $A(\omega_r)$ have positive traces and determinants [see Eqs.~(\ref{RPA_condition})]. It is clear from (\ref{det-A}) that $\det A(\omega_r)$ can be made negative only when at least one of the $\Omega_i$ is complex. Moreover, such $\Omega_i$ must be purely imaginary; otherwise, $\Omega_i^*$ also appears in the product of Eq.~(\ref{det-A}) making the combined contribution from $\Omega_i$ and $\Omega_i^*$ positive. When $\Omega_i$ is purely imaginary, its complex conjugate belongs to the same ``pair'' $\pm\Omega_i$ of eigenvalues of the RPA matrix and does not give a separate contribution to the product in Eq.~(\ref{det-A}). As the temperature decreases and an RPA frequency becomes purely imaginary, the first determinant that becomes singular is the one with the smallest $\omega_r$, i.e., $\omega_1=2\pi T$. Consequently, all determinants $\det A(\omega_r)$ are positive if $|\Omega_i| < 2\pi T$ for all purely imaginary RPA frequencies $\Omega_i$. There is a critical temperature $T_*$ below which this condition is no longer satisfied for some value of the static gap $\Delta_0$.

In principle, the condition $|\Omega_i| < 2\pi T$ is necessary but not sufficient. Simulations show that for a small static field $\Dstaticarg$, the eigenvalues of $A(\omega_r)$ may form a complex-conjugate pair such that $\det A(\omega_r)$ positive, while their real parts may be negative. This can happen at temperatures of the order of or lower than $T_*$.

\end{document}